\documentclass[table]{ccjnl}
\usepackage{lipsum, amsmath}
\usepackage{cuted}
\usepackage{bm}
\usepackage{diagbox}

\graphicspath{{figures/}}
\usepackage{amssymb}
\title{Deep Learning for Multi-Antenna Modulation Recognition of Radio Signals}
 \author{Tao Chen\inst{1,2}, Shilian Zheng\inst{2,*}, Jiepeng Chen\inst{2}, Zhangbin Pei\inst{2}, Qi Xuan\inst{3}, Xiaoniu Yang\inst{2}\corinfo{lianshizheng@126.com}}
\receiveddate{}
\reviseddate{}
\Editor{}

\address[1]{ College of Electrical and Information Engineering, Quzhou University, Quzhou 324000, China}
\address[2]{ National Key Laboratory of Electromagnetic Space Security, Jiaxing 314033, China}
\address[3]{Institute of Cyberspace Security, and also with the College of Information Engineering, Zhejiang University of Technology, Hangzhou 310023, China}
\begin{document}

\maketitle
\begin{abstract}
Multi-antenna receiving systems have become a prevalent technical solution in communication systems. Meanwhile, deep learning has achieved significant progress in automatic modulation recognition tasks in single-antenna systems. However, the application of deep learning in multi-antenna  modulation recognition (MAMR) tasks is still limited. In this paper, we propose an MAMR method namely MAMR-IQ to fully explore the diversity gain of a multi-antenna receiving system, which concatenates the raw received in-phase and quadrature (IQ) signals of  multiple antennas and feeds them into a convolutional neural network. Simulation results show that the proposed MAMR-IQ method outperforms two existing deep learning-based MAMR methods which are based on direct voting (DV) and weight average (WA) in terms of both recognition accuracy and computational complexity. To address the problem of limited training data in few-shot scenarios, we further propose a data augmentation method that involves exchanging IQ sequences received by any two antennas to generate augmented samples. Simulation results show that with the proposed augmentation method, the recognition accuracy can be further improved.
\end{abstract}
\section{Introduction}
Automatic modulation recognition (AMR) has always been a research hotspot in both military and civil applications \cite{123}. In military field, modulation recognition is often used in electronic warfare to help identify the signal of threats \cite{7074190}. In civil field, modulation recognition is used in spectrum management to ensure the security of communications \cite{6852082,2017Cognitive,9020304}. 

The modulation recognition technology with a single receiving antenna has been extensively studied \cite{9339255,8669002,6175327}. These modulation recognition methods basically include two categories, i.e, likelihood-based (LB) methods, and feature-based (FB) methods. LB methods construct a likelihood function of the received signal and then compares it with the predefined threshold to select the modulation type of the output with the maximum probability. According to different prior knowledge, LB methods are divided into average likelihood ratio test (ALRT), generalized likelihood ratio test (GLRT), and hybrid likelihood ratio test (HLRT) \cite{dobre2007survey}. In \cite{4746878}, the authors showed that the ALRT-based method for modulation recognition has higher accuracy than the joint power estimation method based on the higher-order moment of the received signal. In the single-input single-output (SISO) system with a random phase and additive white Gaussian noise (AWGN) channel \cite{904013}, the GLRT-based method achieves significant performance gains over the ALRT-based ones for the classification of non-constant envelope modulations. % The author regards the classification of modulation types and space time block code (STBC) as a joint classification problem, and proposes a joint hybrid likelihood ratio test (J-HLRT) method to find the optimal solution for this joint problem \cite{7750587}. 
Although LB methods can get the optimal solution in theory, the solution of the likelihood function requires perfect prior knowledge and high computational complexity which restrict their application in real-world systems. In contrast, FB methods are computational efficient which mainly include two steps: feature extraction of received signal and the design of classifier \cite{8770611}. The commonly used features include instantaneous features and statistical features. The instantaneous features are usually extracted from instantaneous amplitude or instantaneous frequency of the signal \cite{664294}. For instance, in \cite{8538057}, the instantaneous amplitude and frequency features of minimum shift keying (MSK) and binary phase shift keying (BPSK) signals were extracted, and then support vector machine (SVM) classifier was used to recognize the modulation through these features. The statistical features include high-order moments and cumulants \cite{8782525}. For instance, in \cite{8038046,7776899,644995,893385} the second-order, fourth-order, and sixth-order moments were extracted from the signal sequence for modulation recognition. A major limitation of FB methods is that the designed features directly determine the performance of modulation recognition, and it is often very difficult to design features adapted to that many modulations to be recognized.

% Furthermore, the author in \cite{6486623} considers the prior knowledge of channel state information and studies the cumulants of different orders and their combinations as feature vectors for the modulation recognition of MIMO signals. A feature engineering framework using the fractal self-similarity mechanism of the IQ constellation is designed to identify modulation types in (2*2) MIMO systems, i.e., transmitted by two antennas and received by two antennas \cite{9163011}.

%In \cite{2015Deep}, scholars provides a detailed introduction to deep learning (DL) and look forward to the future of deep learning.
In recent years, deep learning (DL) \cite{2015Deep} has made remarkable achievements in computer vision \cite{9270498}, natural language processing \cite{9885908}, complex networks \cite{8954863} and other fields. DL-based method adopts an end-to-end learning method, i.e., the data is directly input to the neural network without manual feature extraction, and multi-layer hidden layers automatically learn feature maps from the input and then classify or identify it. Because of the above advantages, the DL-based models are applied to modulation recognition in the single-antenna reception system. Initially, deep neural networks (DNNs) are applied in the field of AMR \cite{9762373}. Subsequently, some advanced convolution neural network (CNN) \cite{10042021,9935275}, recurrent neural network (RNN) \cite{9552668,8446021}, and graph neural network (GNN) \cite{9943706} models have been applied to modulation recognition. For example, the authors \cite{2016Convolutional} simulate and generate RML2016.10a dataset with 11 modulation types, and feed in-phase and quadrature (IQ) sequence of the generated dataset to CNN network for recognition. In order to solve the degradation problem of deep learning caused by the increase of neural network layers, deep residual networks are used to signal recognition of dataset with frequency shift and multipath fading \cite{8267032}, and has a greater performance improvement compared with CNN at high SNR. Due to the communication signal is a time series, the three-layer Long Short-Term Memory (LSTM) is used to construct DNN classifier to identify three types of phase shift keying (PSK) and one type of quadrature amplitude modulation (QAM), i.e., BPSK, QPSK, 8PSK and 16QAM in \cite{8891763}. Adaptive Visibility Graph Neural Network (AvgNet) \cite{9695244} is designed to perform end-to-end conversion of IQ sequences into graphs, leveraging both the time series characteristics of signals and the topological structure of graphs for  recognition.

Due to the combination of transmit diversity and receiving antenna arrays in multiple-input multiple-output (MIMO) system, MIMO technology can provide two types of gains, i.e. spatial diversity and spatial multiplexing. %The diversity gain enhances the reliability of the communication link and the multiplexing gain improves the channel capacity. Meanwhile, since the white noise on different antennas is irrelevant, but the noise remains unchanged after the combination, and the energy is multiplied after the combination to obtain array gain  \cite{2017Deep,2013Wideband,2015Compact}.
Due to the popularity of MIMO in current communications systems and networks, researchers have begun to study multi-antenna modulation recognition (MAMR) \cite{8555624,9138373,8600148} and a variety of deep learning-based methods applied to SISO system are gradually applied to MIMO systems \cite{9042355, 9348790,9845419}. For instance, the authors in \cite{98328831} propose a MAMR method based on semi-supervised and deep learning, which combines Generative Adversarial Network (GAN) and one-class Support Vector Machine (1SVM) to identify three digital signals and three analog signals in MIMO system. In \cite{9018261}, MAMR based on CNN network is proposed. The CNN gives the prediction results for the signal received by each receiving antenna, and then the decision-making methods of direct voting (DV), weight voting (WV), direct average (DA) and weight average (WA) are respectively used to give the modulation type of the received signal according to the prediction results. %\textcolor{blue}{The authors \cite{9845419} address the issue of AMR over channels with low rank and employ a combination of feature fusion and deep learning-based classifiers to detect the crucial signal within the transmission. This effectively resolves the AMC challenges presented by rank-deficient channels.} 
However, these four methods utilize neural networks to identify signals received by individual antennas, neglecting the interdependence among multiple antennas in the learning process, which may lead to a decrease in performance. Moreover, processing the received signal sequence of each antenna through CNN raises the complexity. To mitigate these problems, in this paper, we propose a MAMR-IQ method which concatenates the raw IQ components of signals received by multiple antennas and then feeds them to our designed neural network to learn the hidden features and thus identify the modulation types. The MAMR-IQ method maximizes the opportunity of learning the correlation among signals received by different antennas and it is expected to improve the recognition performance. % utilization of IQ components received by multiple antennas, thereby improving system performance while taking into account the interdependence among multiple antennas, and  minimizing time and space complexity as well. 
%In a multi-antenna receiving system, after the signals received by multiple antennas are combined, the probability of simultaneous occurrence of deep fading on different antennas is reduced to obtain diversity gain. 

% The performance of deep learning-based algorithms strongly relies on a large amount of training data. In practice, the training samples need to be collected over-the-air by the receiving system. However, collecting enough radio data in advance is almost impossible in a non-cooperative communication scenario. The AMR under the few-shot conditions seems a significant topic. Recently, more and more attention has been drawn to few-shot learning in the field of AMR 
Deep learning algorithms depend heavily on having a substantial amount of training data. In practical situations, the training samples must be gathered wirelessly by the receiving system. However, it is extremely challenging to collect a sufficient amount of radio data in advance, especially in non-cooperative communication scenarios. Consequently, few-shot learning in the context of automatic modulation recognition (AMR) has emerged as a crucial research area. Recently, there has been a growing interest in few-shot learning for AMR \cite{9245503,9650842}. The early augmentation of modulation recognition is based on classic signal processing, including rotation, flipping, adding Gaussian noise, and time-frequency domain transformation. Specifically, in \cite{8936957}, the authors applied three data augmentation methods to IQ sequence and then fed these data into the neural network for training. In \cite{9858310}, the fast Fourier transform (FFT) time-frequency transform was applied in data augmentation. The training set is extended by the FFT transform of the raw IQ sequence. Secondly, deep learning is also used to generate augmented samples. For example, GAN is used as a tool area for data augmentation to solve the problem of insufficient training samples \cite{9531296}. The authors in \cite{8319926} use the auxiliary classifier generative universal networks (ACGANs) as a data augmentation generator to expand the dataset and improve the performance gain. However, all of these augmentation methods are designed for single-antenna AMR. %require a sufficient amount of data sets to broaden the sample space. As the augmentation procedure is executed, a substantial amount of irrelevant data may be produced, leading to overfitting of the model during the training, consequently diminishing the classification performance. 
%Limited training samples can significantly impact the performance of modulation recognition. Due to the poor accuracy of the neural network caused by insufficient samples,
In this paper, we propose a data augmentation method for MAMR to enlarge the training set and thus improve the recognition accuracy. Our augmentation method involves exchanging IQ sequences received by any two antennas to obtain an augmented sample. %This method takes into account the practical scenario of swapping positions within the antenna array for the purpose of augmentation. %By generating these samples, the signal characteristics are expanded while avoiding the overfitting phenomenon that can occur in neural networks. 
It can be easily combined with existing augmentation methods such as flipping which can further expand the dataset and achieve performance gains.

In summary, the main contributions of this paper are as follows.
\begin{itemize}
\item To fully explore the diversity gain of a multi-antenna receiving system, we propose a MAMR-IQ method which concatenates the raw IQ components of signals received by multiple antennas and then feeds them to our designed convolutional neural network to learn the hidden features and thus identify the modulation types. Compared with existing deep learning-based methods, MAMR-DV and MAMR-WA, our proposed MAMR-IQ has higher recognition accuracy, which demonstrates the superiority of our proposed method.
%\item We evaluate the effectiveness of our proposed MAMR-IQ method in the presence of phase offset, and the simulation results are consistent with the experimental results obtained without phase shifts. The results indicate that the MAMR-IQ method outperforms MAMR-Single, MAMR-DV and MAMR-WA methods.
\item We propose a data augmentation method which exchanges IQ sequences received by any two antennas to enlarge the training set. This operation can be taken several times for a single sample with corresponding label, and then, the augmented sample set can be obtained to improve the recognition accuracy. %The simulation results demonstrate a significant improvement in recognition accuracy with the proposed augmentation method. 
Furthermore, our proposed method can be easily combined with existing flipping augmentation methods which enables us to expand the dataset and achieve further performance gains.
\item We conduct a comparison of the complexity with those of MAMR-DV, MAMR-WA, and MAMR-GAN methods. Our findings indicate that MAMR-IQ exhibits significantly lower time and space complexity compared to MAMR-DV and MAMR-WA methods, which shows its advantage in deployment in real-world systems.
\end{itemize}

The rest of this paper is organized as follows. In Section \uppercase\expandafter{\romannumeral2}, we discuss the received signal 
 mathematical model and the MAMR task. In Section \uppercase\expandafter{\romannumeral3}, the proposed MAMR-IQ and data augmentation methods are introduced. In Section \uppercase\expandafter{\romannumeral4}, we give the simulation results and discuss the complexity of different methods. Finally, Section \uppercase\expandafter{\romannumeral5} concludes the paper.

%----------------------------------------------------------------------------------------
\section{System Model}
\label{s2}
\subsection{Signal Model}
\label{s2-1}
%添加序号前面A,B小标题
%\subsection{Mathematical Model of DSSS Signal Detection}
The modulated signal sent by the transmitter is  received by multiple antennas through wireless transmission \cite{9124713, 8082745}, which can be expressed as
\begin{equation}
\begin{aligned}
\textbf{y}(n)= \textbf{g}(n) x(n)+\textbf{w}(n), n=0,1,\dots,N-1,
\end{aligned}
\label{rt}
\end{equation}
where $\textbf{y}(n)=\left [y_1(n), y_2(n), \cdots, y_C(n) \right]^{T}$ represents the vector of the received signal, $y_i(n)$ is the received signal of the $i$-th antenna, $x(n)$ is the transmitted signal, $C$ is the number of receiving antennas, $\textbf{g}(n)$ is the vector of channel responses, $\textbf{w}(n)$ is the vector of additive white Gaussian noise (AWGN) \cite{4814291}, and $N$ is the legnth of the signal. $\textbf{g}(n)$ and $\textbf{w}(n)$ can be further expressed as
\begin{equation}
\begin{aligned}
\textbf{g}(n)= \left [g_1(n), g_2(n), \cdots,   g_C(n)  \right]^{T},
\end{aligned}
\label{gt}
\end{equation}
\begin{equation}
\begin{aligned}
\textbf{w}(n)= \left [w_1(n), w_2(n), \cdots,   w_C(n)  \right]^{T},
\end{aligned}
\label{wn}
\end{equation}
where $g_i(n)$ and $w_i(n)$ are the channel gain and noise of the $i$-th antenna, respectively.
% For the sake of further computation, the signal received by the $i$-th antenna can be represented as
% \begin{equation}
% \label{ri}
% r_{i}(n)=\left[\begin{array}{c}
%  I_i(n)\\
% Q_i(n)
% \end{array}\right]=\left[\begin{array}{c}
%  \text{real}(y_i(n))\\
% \text{imag}(y_i(n))
% \end{array}\right],
% \end{equation}
% where $\text{real}(\cdot)$ and $\text{imag}(\cdot)$ represent the real part and imaginary part of the signal received by the $i$-th antenna, $I_i(n)$ and $Q_i(n)$ are the in-phase (I) and quadrature (Q) components of the signal $y_i(n)$.

\subsection{MAMR Task }
\label{s2-2}
MAMR is to find out the modulation type of the signal from the set of candidate modulation types according to the received multi-antenna signals. MAMR can be regarded as a multi-category classification problem which can be expressed as
\begin{equation}
\begin{aligned}
Y_i=\underset{m \in\{1,2, \ldots, M\}}{\arg \max }\rm Pr \it (Y_i=m \mid \textbf{y}(n) ),
\end{aligned}
\label{ri1}
\end{equation}
where $\rm Pr(\cdot)$ is to calculate the probability that the modulation type of the received signal belongs to the candidate set of modulation types, $M$ is the number of modulation types. In this paper, $M$ is equal to 12, meaning that the number of signal modulation types used is 12, including four types of PSK, such as BPSK, 8PSK, QPSK, OQPSK, three types of frequency shift keying (FSK), namely 2FSK, 4FSK, 8FSK,  three types of QAM, 16QAM, 32QAM, 64QAM, as well as 2 types of pulse amplitude modulation (PAM), 4PAM, and 8PAM. Thus the classification problem can be realized by deep learning as discussed in the next section.
%$\textbf{y}(n)$ is the input of the neural network, $Y_i$ is the output of the neural network.

\section{Proposed Method}
\label{s3}
\subsection{Input Format}
\label{s3-1}
\begin{figure}[t]
\centering
\includegraphics[width=0.5\textwidth]{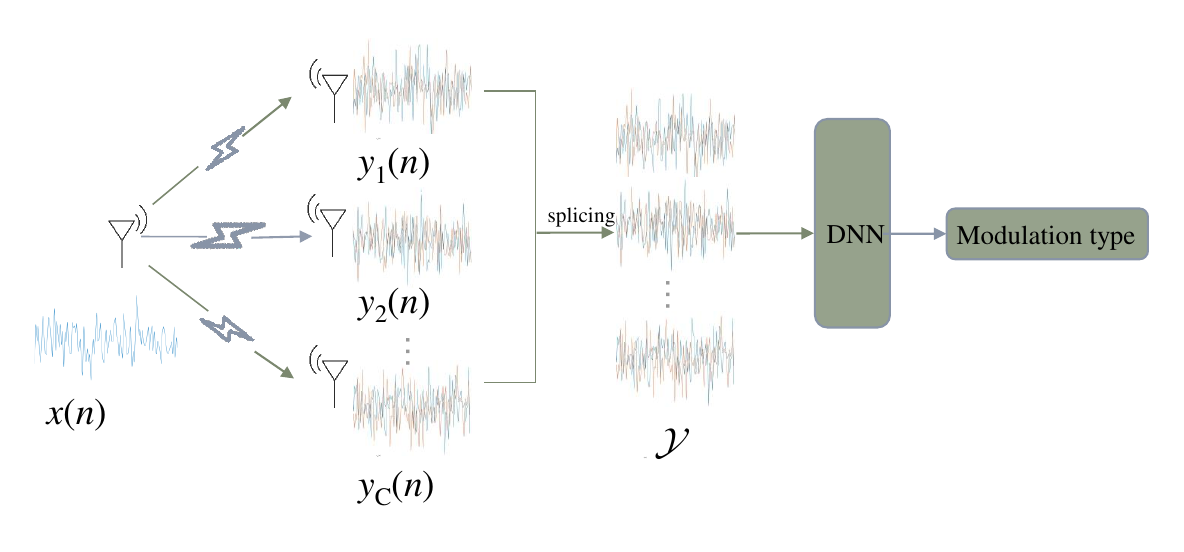}
\caption{The process of modulation recognition of multi antenna receiving system based on Deep Learning.}
\label{fi1}
\end{figure}

The overall process of the proposed MAMRnet is shown in Figure. \ref{fi1}. Firstly, $C$ signals are received through $C$ antennas for the transmitted signal $x(n)$, i.e. from $y_1(n)$ to $y_C(n)$. For the sake of further computation, we extract the IQ components of the signal $y_{i}(n)$ received by the $i$-th antenna by 
\begin{equation}
\label{ri}
\left[\begin{array}{c}
 I_i(n)\\
Q_i(n)
\end{array}\right]=\left[\begin{array}{c}
 \text{real}(y_i(n))\\
\text{imag}(y_i(n))
\end{array}\right],
\end{equation}
where $\text{real}(\cdot)$ and $\text{imag}(\cdot)$ represent extracting the real part and imaginary part of the signal received by the $i$-th antenna, $I_i(n)$ and $Q_i(n)$ are the in-phase (I) and quadrature (Q) components of the signal $y_i(n)$. Afterwards, the IQ components of the received signals are spliced to obtain $\mathcal{Y}$ as

\begin{equation}
\label{M}
\mathcal{Y} = \left[\begin{array}{cccc}
I_{1}(0) & I_{1}(1) & \cdots & I_{1}(N-1) \\
Q_{1}(0) & Q_{1}(1) & \cdots & Q_{1}(N-1) \\
I_{2}(0) & I_{2}(1) & \cdots & I_{2}(N-1) \\
Q_{2}(0) & Q_{2}(1) & \cdots & Q_{2}(N-1) \\
\cdots & \cdots & \cdots & \cdots \\
I_{C}(0) & I_{C}(1) & \cdots & I_{C}(N-1) \\
Q_{C}(0) & Q_{C}(1) & \cdots & Q_{C}(N-1) \\
\end{array}\right] .
\end{equation}

Finally, the spliced $\mathcal{Y}$ are fed into the DNNs designed to obtain the modulation type of the transmitted signal $x(n)$. Due to the powerful feature learning ability of deep learning, this processing method of directly feeding signals from multiple antennas into neural networks will be more conducive to learning the signal correlation between each antenna and is expected to improve recognition performance.

% In the training process, we use backpropagation and cross-entropy as the loss function
% \begin{equation}
% \label{loss}
% \mathcal{L}=-\frac{1}{B}\sum_{j=1}^{B} \sum_{m=1}^{M} \rm Pr\left(\it Y_{jm}\right) \log \left(\rm Pr\left(\it \hat{Y_{jm}}\right)\right),
% \end{equation} 
% where $B$ represents the mini-batch size, $\rm Pr\left(\it Y_{jm}\right)$ is the correct probability that the modulation type of the $j$-th signal sequence is $m$, $\rm Pr\left(\it \hat{Y_{jm}}\right)$ is the predicted probability that the modulation type of the $j$-th signal sequence is $m$.

\subsection{Adopted Neural Network}
\label{s3-2}

\begin{figure}[t]
\centering
\includegraphics[width=0.25\textwidth]{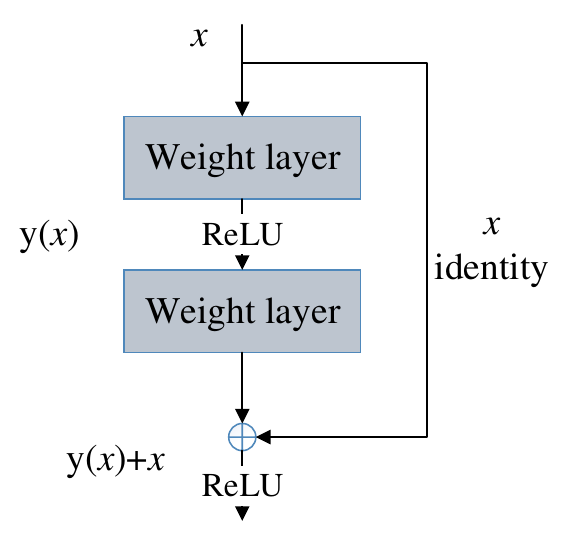}
\caption{The structure of residual.}
\label{residual}
\end{figure}

CNNs are a type of deep learning model specifically designed for processing grid-structured data. They have achieved significant accomplishments in the field of computer vision and are gradually being applied in signal processing to address corresponding issues \cite{8476607,9174643}.
%Generally, a CNN is composed of an input layer, an output layer and multiple hidden layers. The hidden layers are usually composed of  convolutional layers, pooling layers and fully connected layers. Nonlinear activation layers are also added to increase the nonlinear expression ability of the network \cite{9027643,10171167}.}
The depth of the network is crucial to the performance of the neural network models. When the depth of the network increases, the network can extract more complex features, such as Alexnet \cite{9498686} and VGG \cite{Simonyan2014VeryDC}, which improve the performance of the model through the stacking of neural networks. However, when the depth increases to a certain extent, the network has a degradation problem, i.e., the performance reaches saturation and even the accuracy decreases. In order to avoid the degradation of deep network, a new network structure ResNet has been proposed. ResNet introduces the residual structure, that is, an identity mapping is added to speed up the training speed and accuracy, without increasing the additional calculation of the network. The residual structure is realized by short circuit connection, which is shown in Figure. \ref{residual}.

\begin{figure}[!ht]
	% \vspace{-2.0em}
\centering
\includegraphics[width=0.35\textwidth]{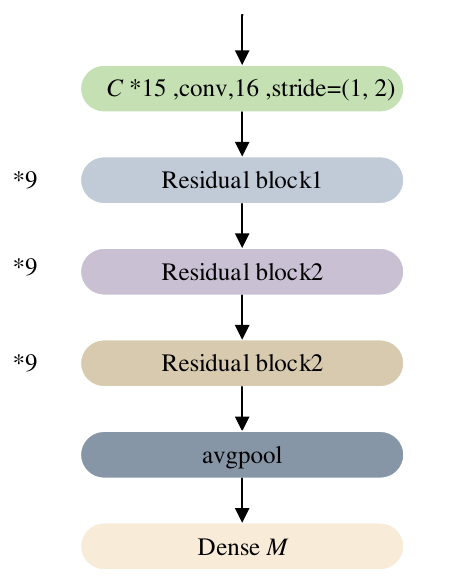}
\caption{Structure of the constructed ResNet56, ``avgpool'' stands for average pooling layer. The repetition times of Residual block1, Residual block2 and Residual block3 are 9.}
\label{resnet}
\end{figure}

We design a ResNet network structure based on the above residual blocks for MAMR in Figure. \ref{resnet}. 
Since the number of layers of ResNet we designed is 56, we denote it as ResNet56. The convolution kernel size of the first convolution layer, $(C,15)$, is designed to adapt to the input size. After the operation, the first dimension of the output feature map is one and one-dimensional convolution can be used in the following layers. There are three residual blocks in the ResNet56, i.e., Residual block1, Residual block2 and Residual block3. The repetition times of the subsequent three residual blocks are 9. The convolution kernel size of the convolution in the three residual blocks is (1, 3), the number of convolution kernels is 16, 32, and 64, and the stride is (1, 1), (1, 2), and (1, 1), respectively. Global average pooling is used to adapt to the  various input lengths. The number of neurons in the last fully connected layer is equal to the number of modulation types, say $M$ in this paper.

\begin{figure*}[htbp]
\centering
\includegraphics[width=0.8\textwidth]{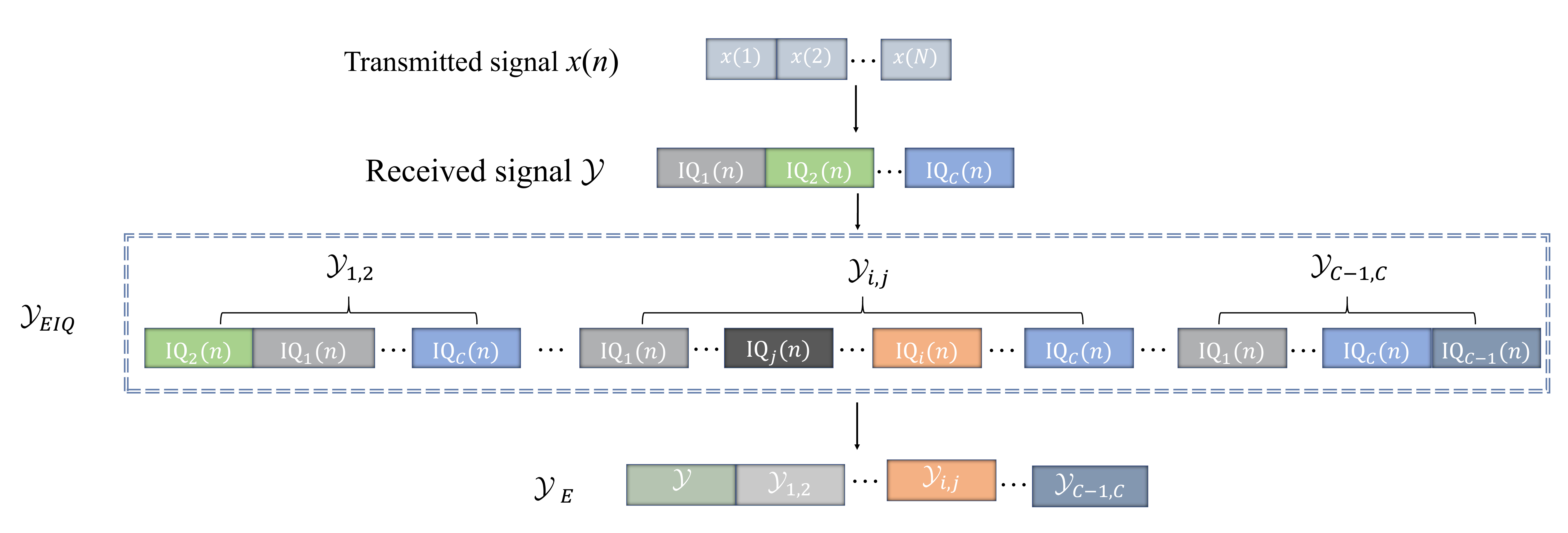}
\caption{The process of data augmentation by exchanging IQ sequences of each antenna, $n=0,1,\dots,N-1$, $N$ represents the length of the signal. $\mathcal{Y}_ {i,j}$ represents the data obtained by exchanging the IQ sequence of the $i$-th antenna and the $j$-th antenna. $\operatorname{IQ}_i(n)$ represents the I-channel and Q-channel components received by the $i$-th antenna.}
\label{exchange2}
\end{figure*}
\subsection{Proposed augmentation Method}
\label{s3-3}
In few-shot scenarios, the number of training samples is limited and the modulation recognition performance will be greatly affected due to insufficient training samples. In view of the poor recognition accuracy of the neural network due to the lack of samples, we propose a data augmentation method in multi-antenna reception to enlarge the training set and thus improve the recognition accuracy. The specific operation process is shown in Fig. \ref{exchange2}. Firstly, the transmission signal with length $N$ is received by $C$ antennas to obtain $\mathcal{Y}$. Then, the IQ sequences received by any two antennas are exchanged to obtain an augmented sample. For example, the IQ sequences received by the $i$-th antenna and the $j$-th antenna are placed as
\begin{equation}
\label{rij}
\mathcal{Y}=\left[\begin{array}{cccc}
\cdots & \cdots & \cdots & \cdots \\
I_{i}(0) & I_{i}(1) & \cdots & I_{i}(N-1) \\
Q_{i}(0) & Q_{i}(1) & \cdots & Q_{i}(N-1) \\
\cdots & \cdots & \cdots & \cdots \\
I_{j}(0) & I_{j}(1) & \cdots & I_{j}(N-1) \\
Q_{j}(0) & Q_{j}(1) & \cdots & Q_{j}(N-1) \\
\cdots & \cdots & \cdots & \cdots \\
\end{array}\right].
\end{equation}
The two sequences can be exchanged to obtain the augmented sample $\mathcal{Y}_{i,j}$ as
\begin{equation}
\label{rij1}
\mathcal{Y}_{i,j}=\left[\begin{array}{cccc}
\cdots & \cdots & \cdots & \cdots \\
I_{j}(0) & I_{j}(1) & \cdots & I_{j}(N-1) \\
Q_{j}(0) & Q_{j}(1) & \cdots & Q_{j}(N-1) \\
\cdots & \cdots & \cdots & \cdots \\
I_{i}(0) & I_{i}(1) & \cdots & I_{i}(N-1) \\
Q_{i}(0) & Q_{i}(1) & \cdots & Q_{i}(N-1) \\
\cdots & \cdots & \cdots & \cdots \\
\end{array}\right].
\end{equation}

This operation can be taken several times for a single sample $\mathcal{Y}$ with corresponding label $\ell$, and then, the augmented sample set can be obtained:
\begin{equation}
\label{reiq}
\mathcal{Y}_{\operatorname{EIQ}}=\{ (\mathcal{Y}_{1,2}, \ell),\ldots,(\mathcal{Y}_{i,j},\ell),\ldots,(\mathcal{Y}_{C-1,C},\ell)\}.
\end{equation}
The maximum number of samples in this set is equal to the maximum number of times we can exchange IQ sequences for $C$ antennas, which can be calculated as
\begin{equation}
\setlength{\abovedisplayskip}{3pt}
\setlength{\belowdisplayskip}{3pt}
D_{\rm{max}}=\frac{C(C-1)}{2}. 
\end{equation}

Finally, we can combine the raw sample $\mathcal{Y}$ received by the $C$ antennas with the augmented sample dataset $\mathcal{Y}_{\operatorname{EIQ}}$ to obtain the final expanded training set of sample $\mathcal{Y}$:
\begin{equation}
\label{re}
\mathcal{Y}_{\mathrm{E}}=\left\{(\mathcal{Y}, \ell), \mathcal{Y}_{\mathrm{EIQ}}\right\}.
\end{equation}

Our proposed data augmentation method can be easily combined with existing augmentation method designed for single antenna signals to further enlarge the dataset. For instance, we can combine the method with flipping-based methods. We flip raw sample $\mathcal{Y}$ received by $C$ antennas in the horizontal, vertical and both directions respectively to obtain flipped samples.
The horizontal flip is to change the value of I channel into its opposite to get obtain the horizontally flipped sample
\begin{equation}
\label{rij11}
\mathcal{Y}_{\operatorname{flipI}}=\left[\begin{array}{cccc}
-I_{1}(0) & -I_{1}(1) & \cdots & -I_{1}(N-1) \\
Q_{1}(0) & Q_{1}(1) & \cdots & Q_{1}(N-1) \\

\cdots & \cdots & \cdots & \cdots \\
-I_{i}(0) & -I_{i}(1) & \cdots & -I_{i}(N-1) \\
Q_{i}(0) & Q_{i}(1) & \cdots & Q_{i}(N-1) \\
\cdots & \cdots & \cdots & \cdots \\
-I_{C}(0) & -I_{C}(1) & \cdots & -I_{C}(N-1) \\
Q_{C}(0) & Q_{C}(1) & \cdots & Q_{C}(N-1) \\
\end{array}\right].
\end{equation}
% Based on this, we can obtain the horizontally flipped dataset 
% \begin{equation}
% \label{flipI}
% \mathcal{Y}_{\operatorname{flipI}}=\{(\mathcal{Y},\ell),(\mathcal{Y}_{\operatorname{flipI_0}},\ell)\}.
% \end{equation}
Similarly, we can convert the value of the Q channel to its opposite to get the vertically flipped sample $\mathcal{Y}_{\operatorname{flipQ}}$. Flipping the values of channel I and channel Q together can get a sample $\mathcal{Y}_{\operatorname{flipIQ}}$.
% \begin{equation}
% \label{I_flip}
% y_{\operatorname{flipI}i}(n)=\left[\begin{array}{c}
%  -I_i(n)\\
% Q_i(n)
% \end{array}\right]=\left[\begin{array}{c}
%  -\text{real}(y_i(n)\\
% \text{imag}(y_i(n))
% \end{array}\right].
% \end{equation}
% and the vertical flipped signal is 
% \begin{equation}
% \label{Q_flip}
% y_{\operatorname{flipQ}i}(n)=\left[\begin{array}{c}
%  I_i(n)\\
% -Q_i(n)
% \end{array}\right]=\left[\begin{array}{c}
%  \text{real}(y_i(n)\\
% -\text{imag}(y_i(n))
% \end{array}\right].
% \end{equation}
By flipping horizontally, vertically, and in both directions, we can expand the number of samples from one to four and obtain the set $\mathcal{Y}_{\operatorname{flipALL}}$:
\begin{equation}
\setlength{\abovedisplayskip}{3pt}
\setlength{\belowdisplayskip}{3pt}
\mathcal{Y}_{\operatorname{flipALL}}=\{(\mathcal{Y},\ell),(\mathcal{Y}_{\operatorname{flipI}},\ell),(\mathcal{Y}_{\operatorname{flipQ}},\ell),
(\mathcal{Y}_{\operatorname{flipIQ}},\ell)\}.
\end{equation}
% where $\mathcal{Y}_{\operatorname{flipI}}$ contains raw samples and vertically flipped samples, $\mathcal{Y}_{\operatorname{flipIQ}}$ contains raw samples and horizontally flipped samples, and $\mathcal{Y}_{\operatorname{flipALL}}$ contains three expanded samples.
These operations can be taken for all of the samples in $\mathcal{Y_{\rm E}}$ and we can get $\mathcal{Y}_{\operatorname{E,flipI}}$, $\mathcal{Y}_{\operatorname{E,flipIQ}}$, $\mathcal{Y}_{\operatorname{E,flipALL}}$. Finally, we can use these augmented datasets to train MAMR network to improve the recognition performance in few-shot scenarios.

\begin{table}[t]
\vspace{-2.0em}
\renewcommand\arraystretch{1.2}
\centering
\caption{Details of Datasets}
\label{table2}
\setlength{\tabcolsep}{2mm}{
\begin{tabular}{cc}
\hline\hline
Data Size  & 2*512 \\ \hline
SNR  & \makecell[c]{ $-20$ dB to 30 dB with interval 2 dB}\\ \hline
 Noise & AWGN \\ \hline
Number of antennas & 2,4,8,16 \\ \hline
Number of Training Samples & 156000 \\ \hline
Number of Test Samples & 156000 \\ \hline
%使用makecell进行行内换行
Modulation Types   & \makecell[c]{2FSK,4FSK,8FSK,\\16QAM,32QAM,64QAM,\\4PAM,8PAM,\\BPSK,QPSK,8PSK,OQPASK} \\ 
 \hline\hline
\end{tabular}}
\end{table}

\section{Performance Analysis}
\subsection{Simulation Setting}
\subsubsection{Datasets}
The dataset we used is generated through simulation using MATLAB, and the specific parameters are set in Table \ref{table2}. The length of each signal is 512, and the oversampling ratio is 8, i.e., there are 64 symbols in a signal. The pulse-shaping filter is a root raised-cosine (RRC) filter with 6-symbols truncated length and the roll-off factor ranges from 0.2 to 0.7. The normalized frequency offset, ranging from $-$0.2 to 0.2, is randomly selected relative to the sampling frequency.
%The frequency offset is randomly chosen between $-0.2$ and 0.2.
The number of signal modulation types used is 12, which are BPSK, 8PSK, QPSK, OQPSK, 2FSK, 4FSK, 8FSK, 16QAM, 32QAM, 64QAM, 4PAM, and 8PAM.
In the simulation, we consider two antenna settings. The first is the fixed setting, assuming that the antenna array used in the training stage and the test stage is fixed. Therefore, in this scenario, there is a fixed phase difference in the signals received by different antennas. The second is random setting. In this scenario, we hope that the trained model can adapt to any antenna array with the same number of antennas. In this case, the phase difference between signals received by different antennas is randomly set.

%In order to verify the accuracy of our method in the phase offset scenario, we add the phase offset in the range of [0, 2$\pi$] for each I and Q channel on the generated dataset.

%In few-shot scenario, we select the dataset with phase offset for the experiment. We define the sample ratio as the ratio of the number of samples of each modulation category in the selected training set to the total number of samples of each modulation category in the training set. Take 0.2\%, 1\% and 5\% samples of each modulation category under each SNR as the received signal, i.e., sample ratio = 0.002, 0.01 and 0.05. The raw smaples $\mathcal{Y}$ received by the $C$ antennas or flipped datasets are randomly exchanged to obtain augmented data. At the same time, in order to explore the influence of exchange times on the accuracy of modulation recognition, we conducted different exchange times on the raw samples $\mathcal{Y}$ and flipped dataset $\mathcal{Y}_{\operatorname{flipI}}$, $\mathcal{Y}_{\operatorname{flipIQ}}$, and $\mathcal{Y}_{\operatorname{flipALL}}$. Finally, we use these augmented datasets to train MAMRnet to improve the recognition performance.
%\begin{center}Table 2\end{center}
\subsubsection{Training Environment}
All simulations are carried out on a notebook with Intel Core i9-12900HX CPU and NVIDIA RTX3080Ti GPU. We are based on the PyTorch framework to recognize the modulation type of the data generated by simulation. The specific training parameter settings are shown in Table \ref{table1}.
%\begin{center}Table 2\end{center}
\begin{table}[t]
\renewcommand\arraystretch{1.2}
\centering
\caption{Training Hyperparameters}
\label{table1}
\setlength{\tabcolsep}{11mm}{
\begin{tabular}{cc}
\hline\hline
Optimizer  & Adam \\ \hline
 Epochs & 20 \\ \hline
Bath size & 128 \\ \hline
Initial learning rate & 0.01 \\ \hline
Learning rate decay rate & 10\% \\ \hline
Learning rate decay epoch & 10, 15 \\ 
 \hline\hline
\end{tabular}}
\end{table}

\subsection{ Simulation Results }
\subsubsection{Comparison of Different Methods} 
We compare the accuracy of our proposed MAMR-IQ method with MAMR-DV and MAMR-WA methods proposed in \cite{9018261}, as well as MAMR-GAN method proposed in \cite{9531296}. The setting of the antenna array is fixed and the number of receiving antennas is 4. The performance of the method with single antenna reception, denoted as MAMR-Single, is also provided for comparison. The networks used are CNN5 in \cite{9018261}, MCLDNN in \cite{9106397}, GAN in \cite{877781}, and ResNet56 adopted in this paper. The experimental results are shown in Table \ref{methods}. It can be observed that MAMR-IQ method we designed demonstrates a superiority over other methods no matter which network is used. For example, with ResNet56, the accuracy of MAMR-IQ method is 0.7829, while the accuracies of MAMR-WA method and MAMR-DV method are 0.7063 and 0.6906, respectively, with a performance gain of about 8\%. Similar performance improvements are observed with the CNN5 and MCLDNN network.

\begin{figure}[!t]
\centering
\subfloat[\label{fig:cnn5}CNN5]{\includegraphics[width=.8\linewidth]{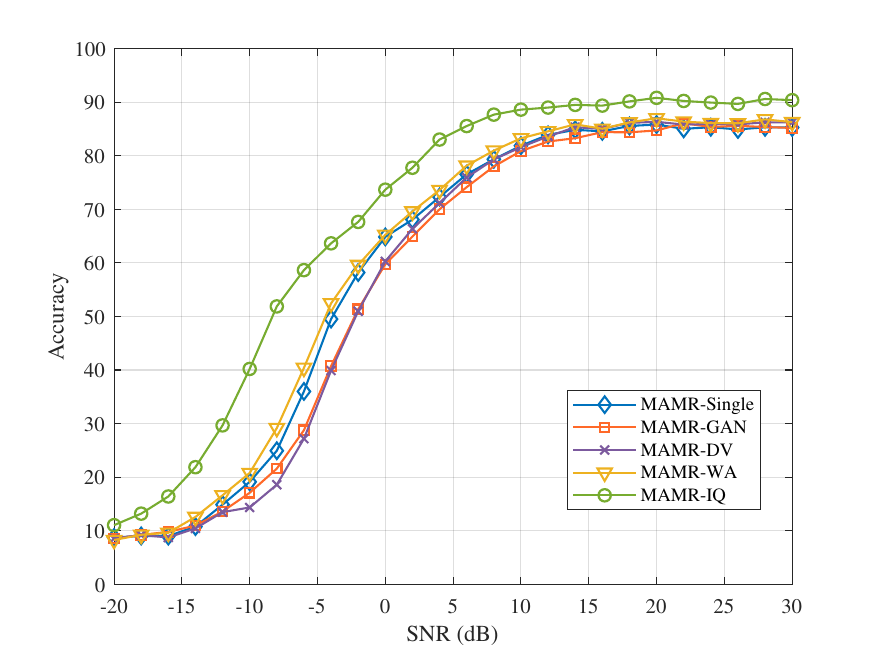}}%
\quad
\subfloat[\label{fig:MCLDNN}MCLDNN]{\includegraphics[width=.8\linewidth]{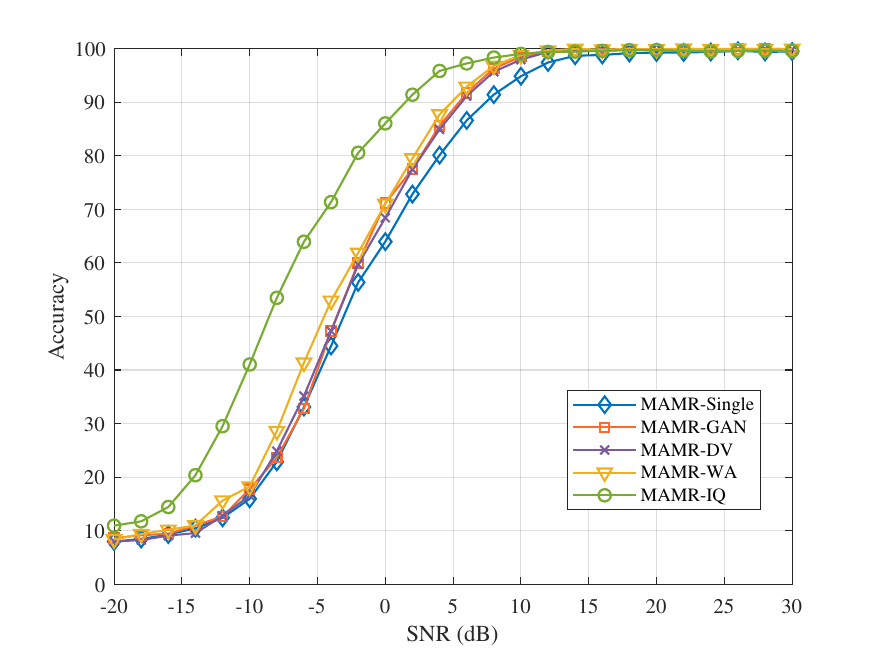}}\\

\subfloat[\label{fig:ResNet56_SNR}ResNet56]{\includegraphics[width=.8\linewidth]{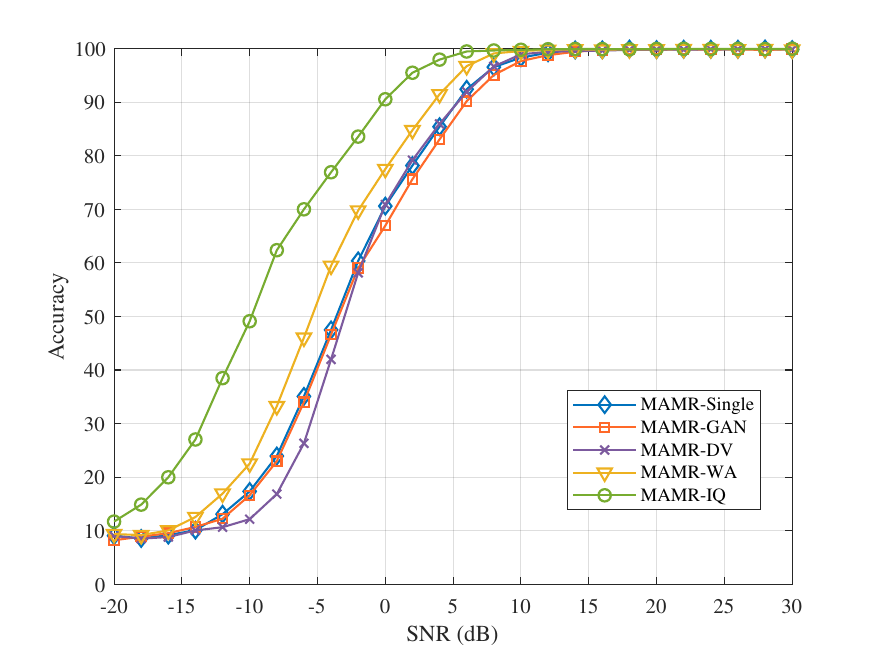}}\\
\caption{Recognition accuracy of different methods.}
\label{snr}
\end{figure}

\begin{figure*}[!ht]
\centering
\subfloat[\label{fig:CNN5-basd MAMR-Single}CNN5-basd MAMR-Single]{\includegraphics[width=.25\linewidth]{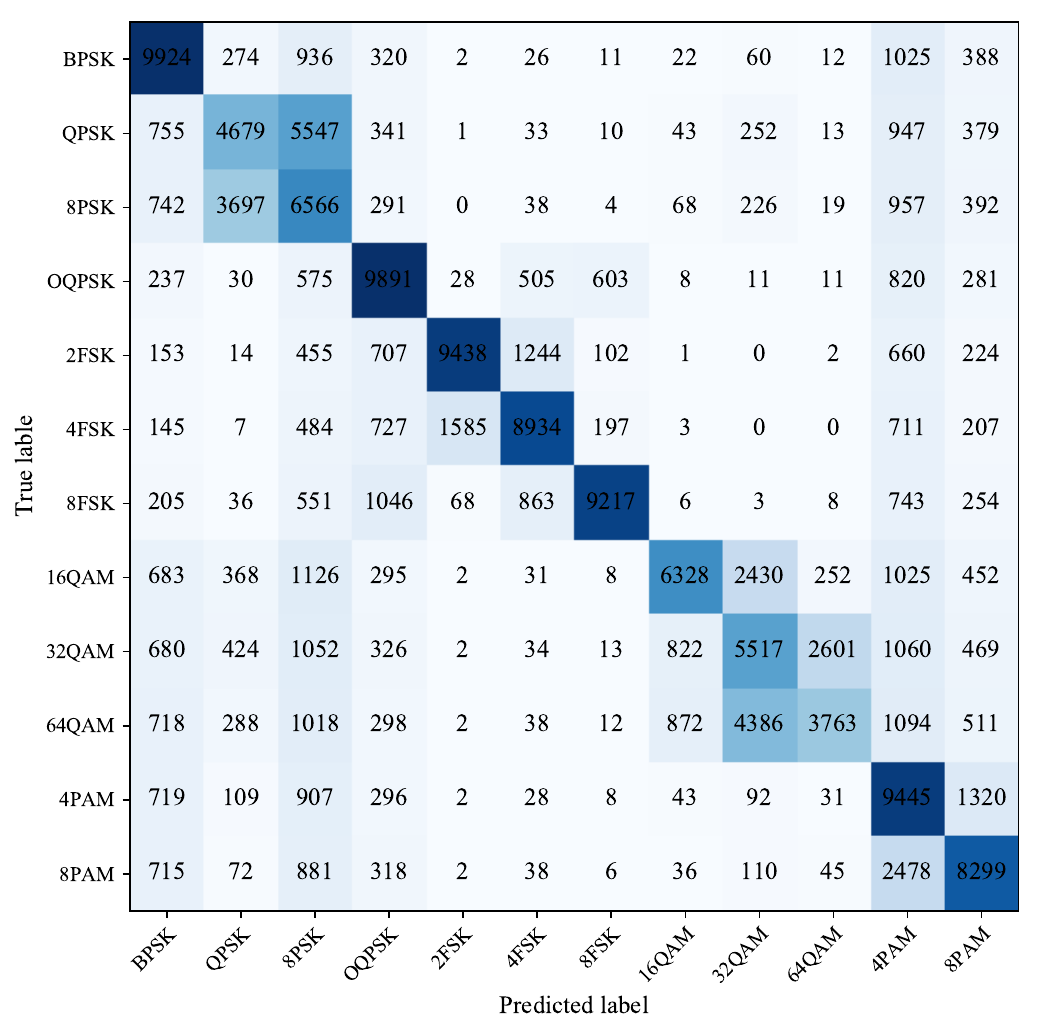}}%
\subfloat[\label{fig:CNN5-based MAMR-DV}CNN5-based MAMR-DV]{\includegraphics[width=.25\linewidth]{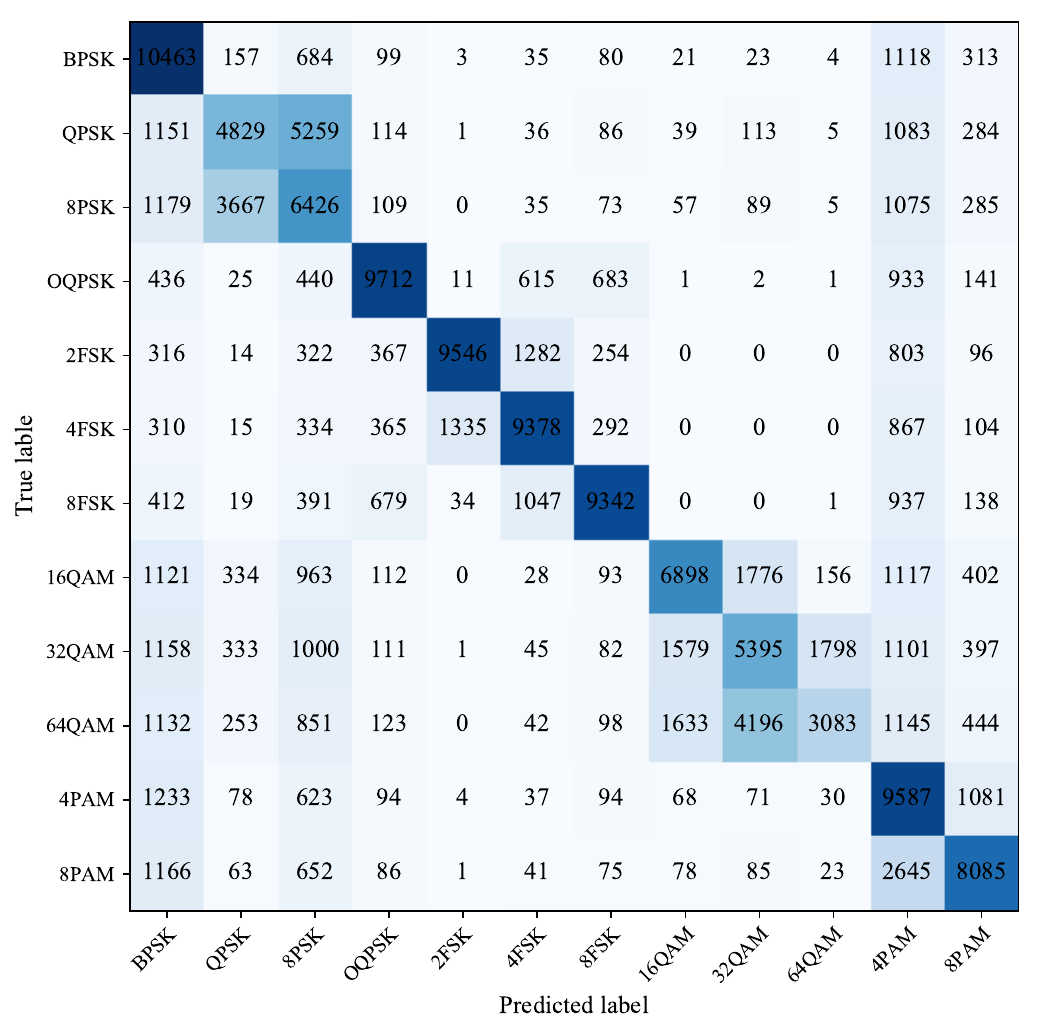}}
\subfloat[\label{fig:CNN5-based MAMR-WA}CNN5-based MAMR-WA]{\includegraphics[width=.25\linewidth]{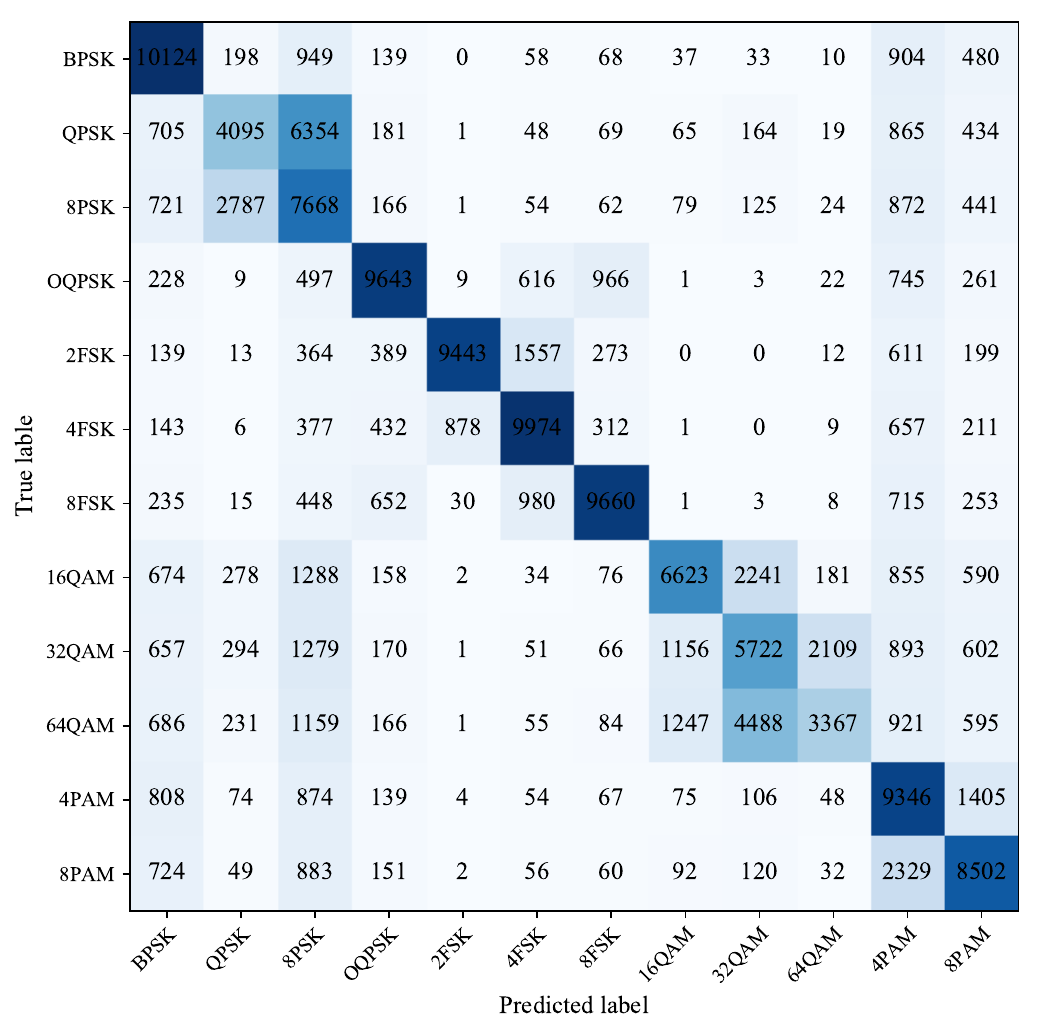}}%
\subfloat[\label{fig:CNN5+WA}CNN5-based MAMR-IQ]{\includegraphics[width=.25\linewidth]{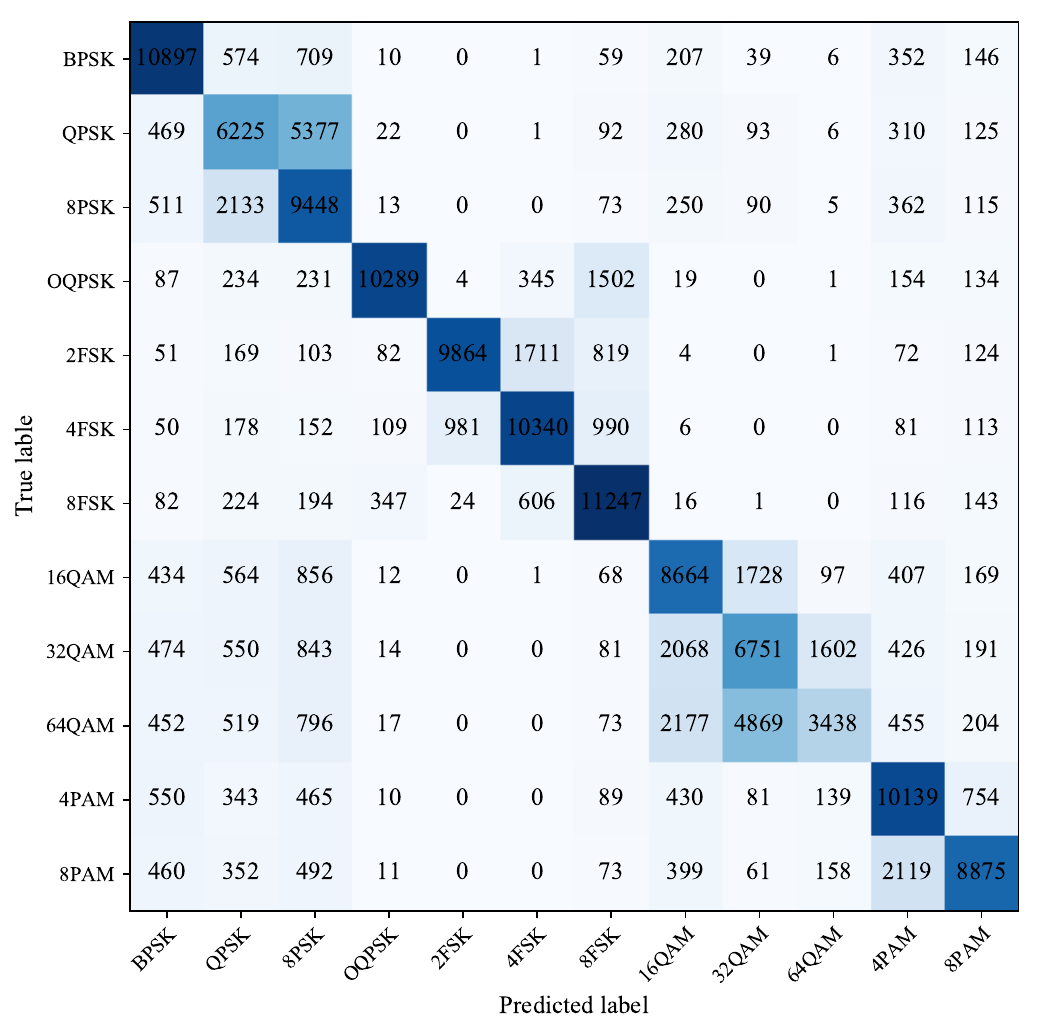}}%

\subfloat[\label{fig:MCLDNN-basd MAMR-Single1}MCLDNN-basd MAMR-Single]{\includegraphics[width=.25\linewidth]{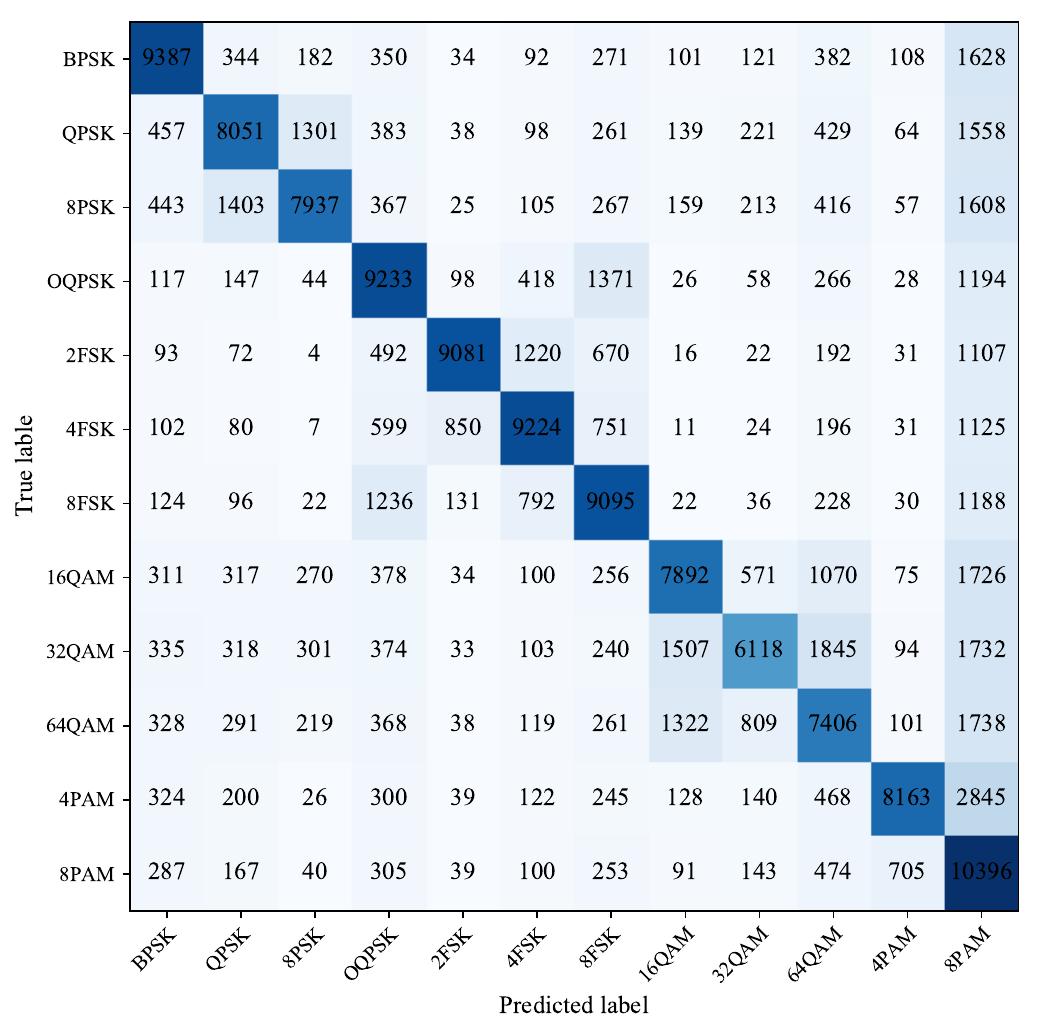}}%
\subfloat[\label{fig:MCLDNN-based MAMR-DV}MCLDNN-based MAMR-DV]{\includegraphics[width=.25\linewidth]{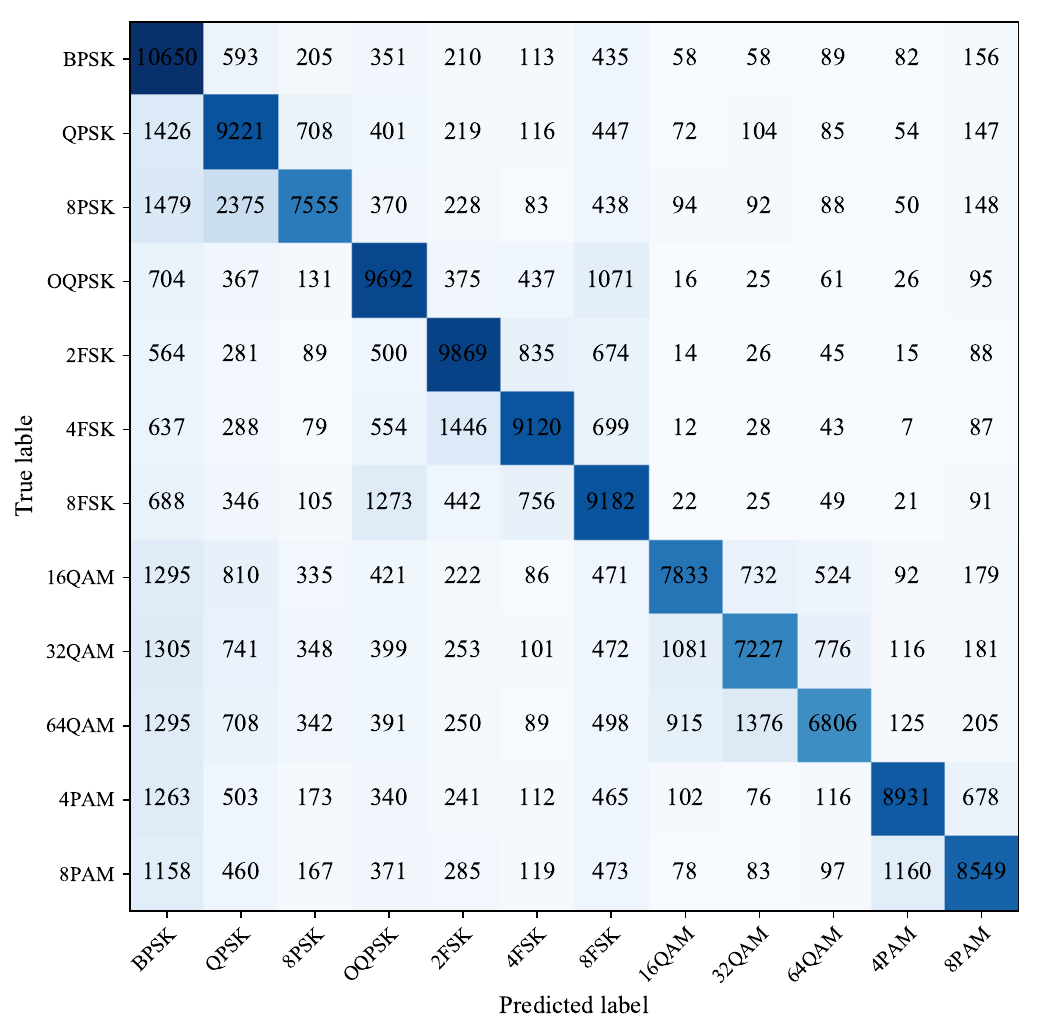}}
\subfloat[\label{fig:MCLDNN-based MAMR-WA}MCLDNN-based MAMR-WA]{\includegraphics[width=.25\linewidth]{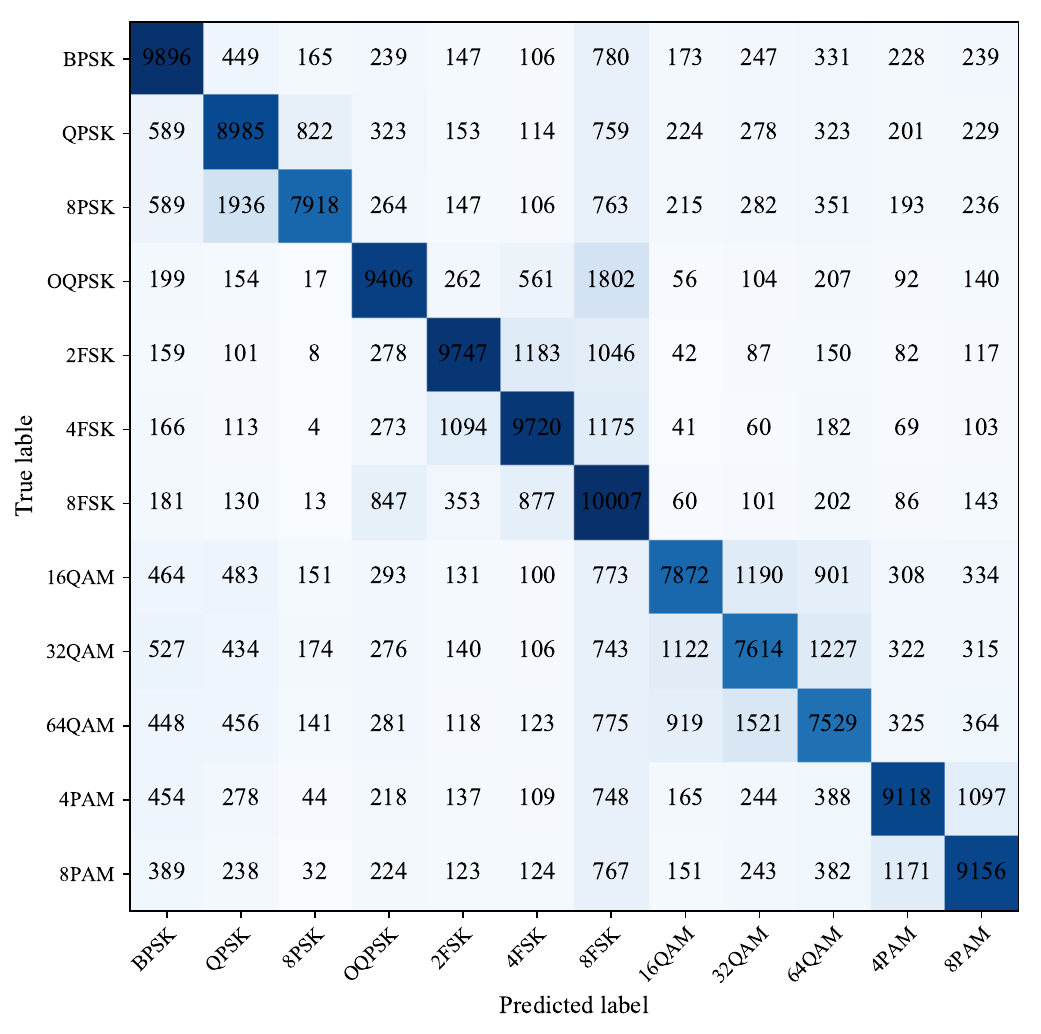}}%
\subfloat[\label{fig:MCLDNN+WA}MCLDNN-based MAMR-IQ]{\includegraphics[width=.25\linewidth]{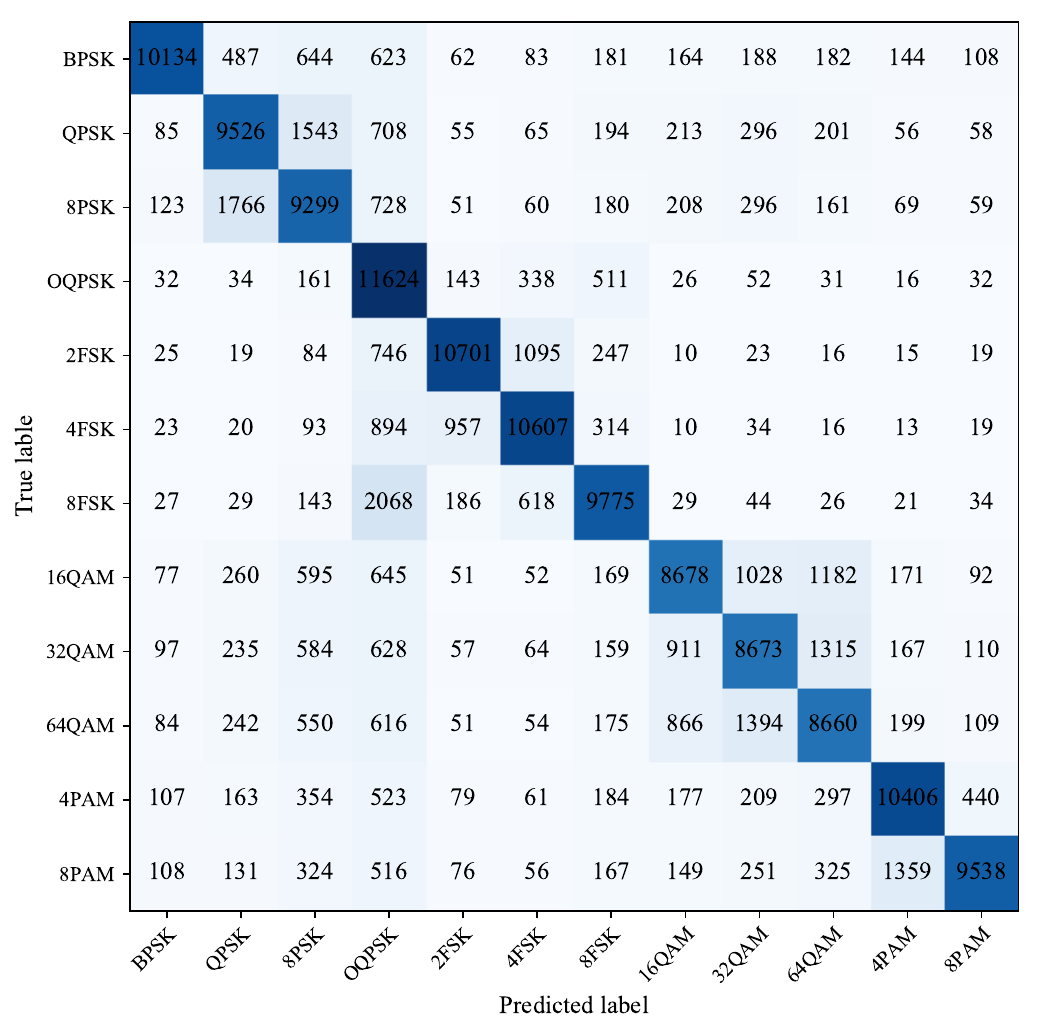}}%

\subfloat[\label{fig:ResNet56-based MAMR-Single} \!ResNet56-based \! \!MAMR-Single]{\includegraphics[width=.25\linewidth]{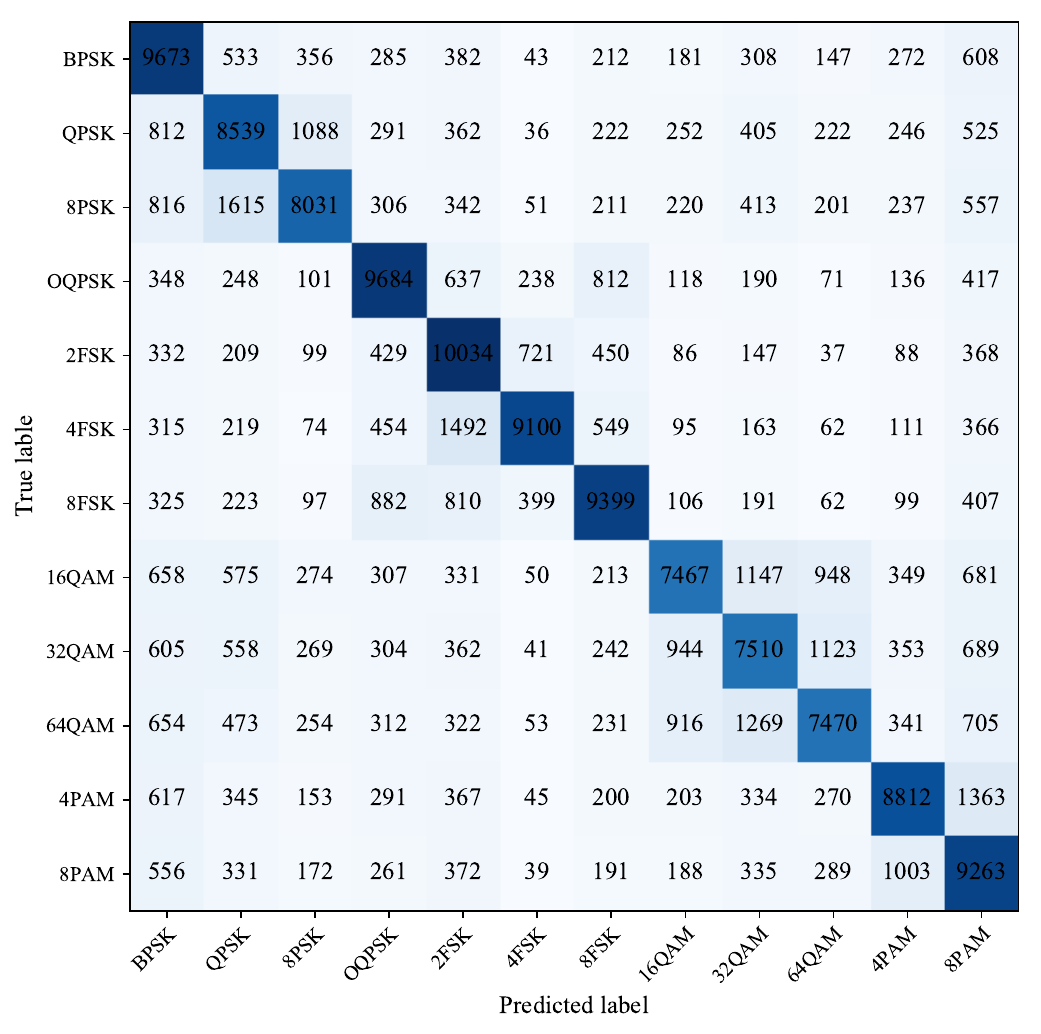}}%
\subfloat[\label{fig:ResNet56-based MAMR-DV}ResNet56-based MAMR-DV]{\includegraphics[width=.25\linewidth]{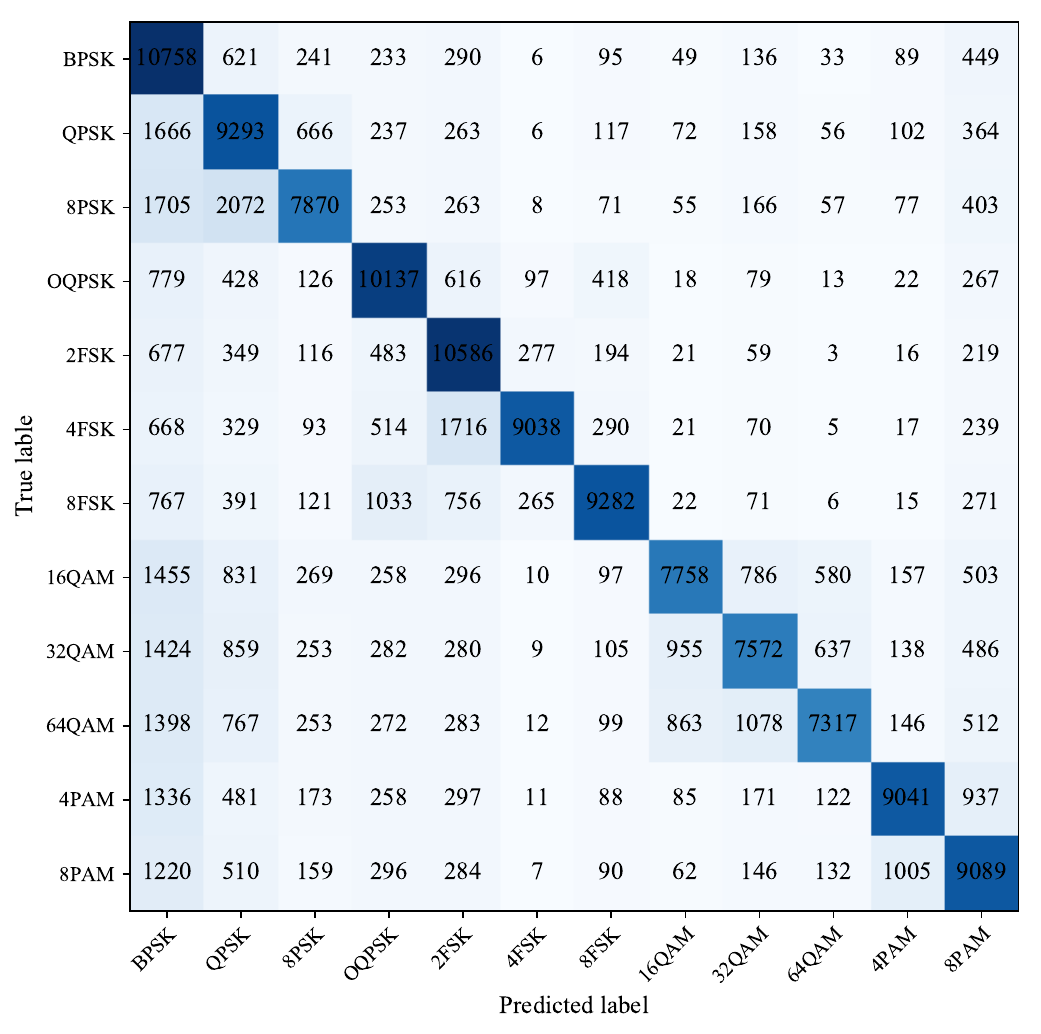}}
\subfloat[\label{fig:ResNet56-based MAMR-WA}ResNet56-based MAMR-WA]{\includegraphics[width=.25\linewidth]{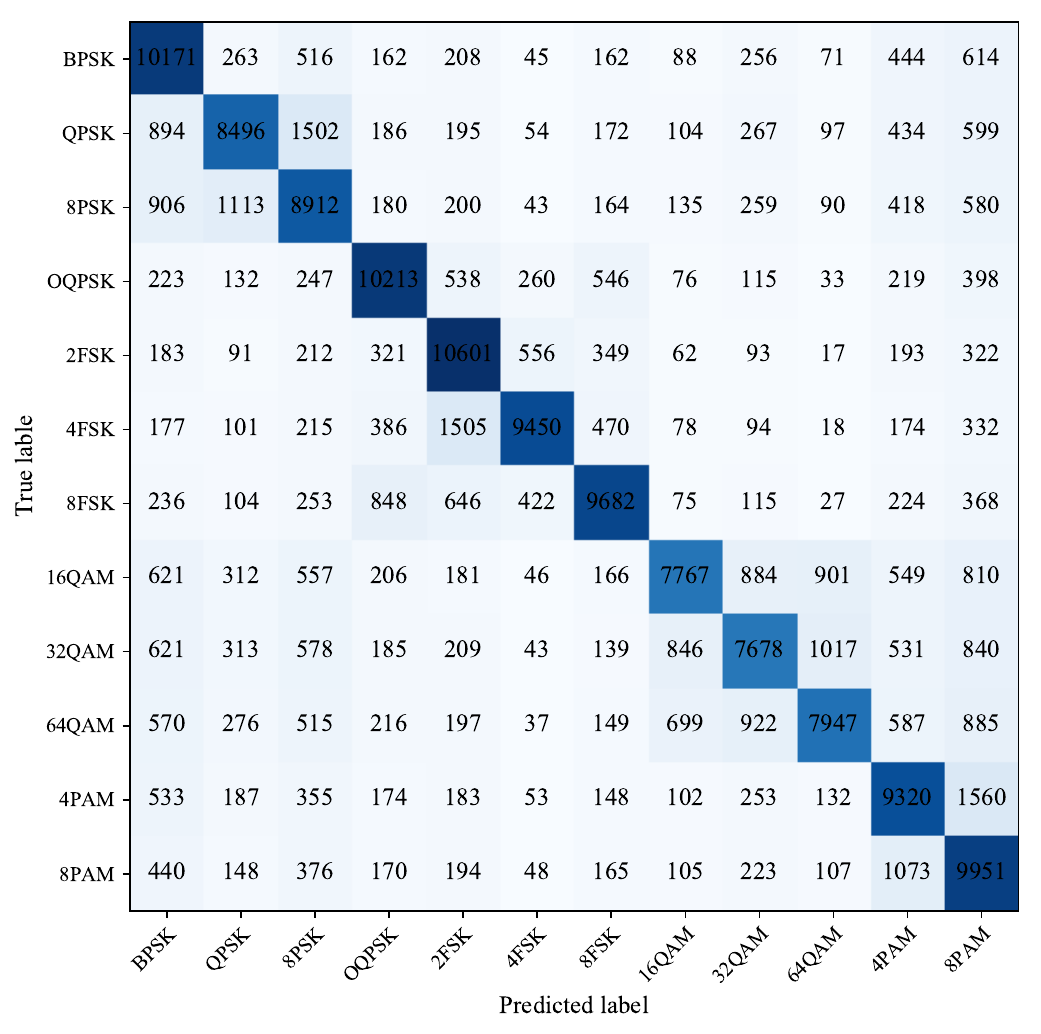}}%
\subfloat[\label{fig:ResNet56+WA}ResNet56-based MAMR-IQ]{\includegraphics[width=.25\linewidth]{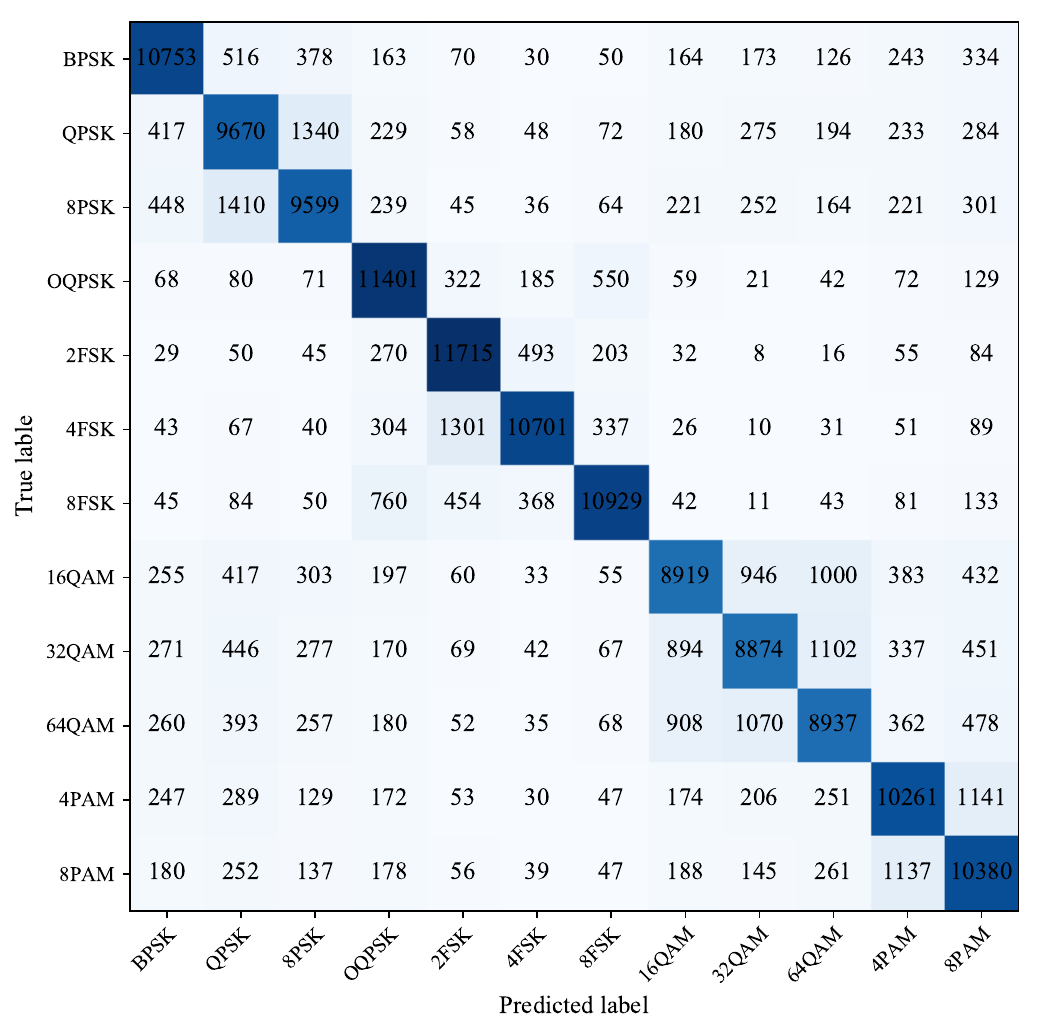}}%
  \caption{ The confusion matrices of different methods.}
\label{input1}
\end{figure*}

%with an improvement of nearly 10\% in classification accuracy. For example, the accuracy of ResNet56-based MAMR-IQ method is 0.7829, while the accuracy of CNN5-based MAMR-IQ method is only 0.681. Additionally, the accuracies of the ResNet56-based MAMR-WA method and  ResNet56-based MAMR-DV method are 0.7063 and 0.6906, respectively. These results underscore the superior performance of MAMR-IQ method compared to the other methods.
We also draw the accuracy of these methods under different SNR in Fig. \ref{snr}. The figure clearly indicates that existing MAMR-DV and MAMR-WA methods do not exhibit significant improvement in terms of accuracy as compared to the MAMAR-Single method. In contrast, the proposed MAMR-IQ method demonstrates significant improvement over the aforementioned three methods across the whole SNR range. Notably, the MAMR-IQ method achieves the same level of accuracy at $-6$ dB as the MAMR-Single method at 0 dB when using the ResNet56 network. This indicates that an SNR gain of about 6 dB has been obtained. However, MAMR-GAN method exhibits minimal improvement, suggesting that GANs may not be suitable for generating high-quality synthetic signal data when compared to their effectiveness in image-related tasks. We also draw the confusion matrices of these methods in Fig. \ref{input1}, excluding MAMR-GAN method. The matrices of MAMR-DV and MAMR-WA methods are almost the same as that of MAMAR-Single method. However, for the modulation schemes QPSK, OQPSK, 4FSK, 8FSK, 16QAM, and 32QAM, our proposed MAMR-IQ method exhibits significant performance gains over MAMR-Single, MAMR-DV, and MAMR-WA methods. Specifically, the number of samples correctly recognized has increased at least 1200 with ResNet56, at least 2000 with CNN5 for each modulation, and at least 800 with MCLDNN, excluding 8PAM. Based on the results of the above experiments, ResNet56 network outperforms CNN5 and MCLDNN, therefore we select ResNet56 network for the subsequent simulation experiments.

  \begin{table}[!t]
\renewcommand\arraystretch{1.3}
\centering
\caption{Performance of different networks}
\label{methods}
\setlength{\tabcolsep}{1.8mm}{
\begin{tabular}{c|ccccc}
\hline\hline
 \diagbox{Network}{Method}&Single&GAN&DV&WA&IQ\\ \hline
CNN5  & 0.5898 &0.5776  & 0.5945 &0.6036 &\pmb{0.6806}   \\ \hline 
MCLDNN  & 0.6638 &0.6655  & 0.6738 &0.6959 &\pmb{0.7156}   \\ \hline 
ResNet56  & 0.6743  &0.6690 & 0.6906 &0.7063 &\pmb{0.7829}   \\ 
\hline\hline
\end{tabular}}
\end{table}

\subsubsection{Effect of Number of Antennas} 
  \begin{table}[t]
\renewcommand\arraystretch{1.3}
\centering
\caption{Performance of different number of antennas}
\label{antenna}
\setlength{\tabcolsep}{4mm}{
\begin{tabular}{c|cccc}
\hline\hline
 \diagbox{$C$}{Method}&DV&WA&IQ\\ \hline
2    & 0.6747 &0.6899 &\pmb{0.7297}   \\ \hline 
4    & 0.6906 &0.7063 &\pmb{0.7829}   \\  \hline 
8    & 0.7081 &0.7223 &\pmb{0.8318}   \\  \hline 
16    & 0.7204 &0.7345 &\pmb{0.8707}   \\ 

\hline\hline
\end{tabular}}
\end{table}
In order to explore the influence of the number of receiving antennas on modulation recognition, we conduct experiments on simulated datasets with varying numbers of receiving antennas. Specifically, we generate datasets with 2, 4, 8, and 16 receiving antennas. The recognition accuracy under different number of antennas is shown in Table \ref{antenna}. As the number of antennas increases from 2 to 16, the accuracy of the MAMR-IQ method also improves continuously. For example, the accuracy of MAMR-IQ method improves from 0.7297 to 0.8707 when the number of antennas increases from 2 to 16. This finding suggests that the recognition accuracy is positively correlated with the number of antennas. Additionally, the performance gain of MAMR-IQ method over MAMR-DV method and MAMR-WA method increases as the number of antennas grows. Specifically, when the number of antennas is 2, MAMR-IQ method outperforms the other methods by 4\%. However, when the number of antennas is 16, MAMR-IQ method exhibits a surprisingly high performance gain of 13.6\%. Overall, MAMR-IQ method we proposed yields superior performance compared to MAMR-DV and MAMR-WA methods.

%%不同天线数目下的准确度图
% \begin{figure}[htbp]
%     \centering
%     \includegraphics[width=7.5cm]{fig/channel.png}
%     \caption{Recognition accuracy under different number of antennas.}
%     \label{channel}
% \end{figure}
\subsubsection{Performance in Random Antenna Setting} 
\begin{figure}[!ht]
\centering
\includegraphics[width=0.4\textwidth]{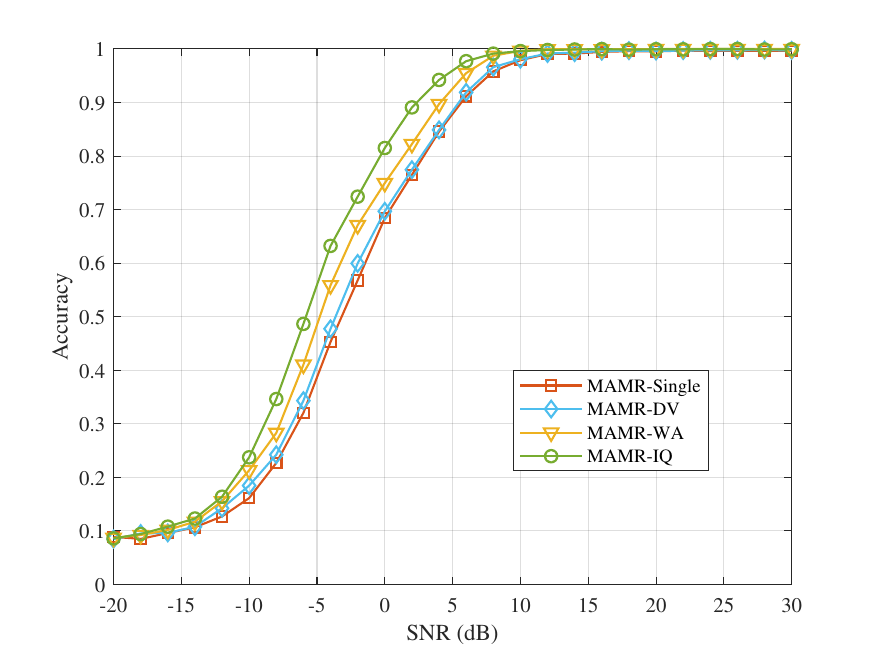}
\caption{Recognition accuracy in random antenna setting.}
\label{xp}
\end{figure}

In real-world scenarios, we may encounter the situation of variant antenna array settings. To account for this, we use the dataset with random phase offset among the received signals of different antenna elements. The number of antennas in this experiment is 2 and the accuracy under different SNR is shown in Fig. \ref{xp}. The experimental results are consistent with the results of the experiment with fixed antenna setting. The recognition accuracy achieved by MAMR-DV and MAMR-WA method is slightly higher than that of MAMR-single method. Furthermore, MAMR-WA method yields better results compared to MAMR-DV method. Our proposed MAMR-IQ method also has superior performance in this scenario. It can be seen from the figure that the accuracy of MAMR-IQ method is higher than that of other methods in the SNR range of $-$15 dB to 10 dB. Similar to the experiment conducted in fixed anttenna setting, an SNR gain about 3 dB over single antenna is  observed in this scenario. %Specifically, MAMR-IQ method achieves the same level of accuracy at $-3$ dB as MAMR-Single method at 0 dB.

% Compared with other methods, MAMR-IQ method also has good performance in the phase offset scenario. It can be seen from the figure that the accuracy of MAMR-IQ method is higher than that of other methods in the SNR range of - 15 dB to 10 dB. The performance gain is the same as the experiment without phase offset. Likewise, the recognition accuracy achieved by MAMR-DV and MAMR-WA method outperforms that of MAMR-single method. Furthermore, MAMR-WA method yields better results compared to MAMR-DV method. Compared with MAMR-Single method, MAMR-IQ method has nearly 3 dB performance gain when the SNR of samples in MAMR-Single method is 0 dB. Notably, the MAMR-IQ method achieves the same level of accuracy at -3 dB as the MAMR-Single method at 0 dB .

% We also draw the confusion matrix of the two methods in Fig. \ref{sample1}. We can see that compared with ResNet56+Each-based method, the recognition accuracy of ResNet56+ALL-based method improved significantly for 8PSK, OQPSK, and 64QAM, and the number of correct recognition increased by more than 1000.

% \begin{figure}[htbp]
%     \centering
%     \includegraphics[width=7.5cm]{fig/ResNet56_SNR_P.pdf}
%     \caption{Recognition accuracy in phase offset scenario.}
%     \label{xp}
% \end{figure}

%相位偏移的混淆矩阵
% \begin{figure}
% \centering
% \subfigure[]{\label{fig1:subfig:a}
% \includegraphics[width=0.45\linewidth]{fig/each_P.pdf}}
% \hspace{-1mm}
% %\hspace{0.01\linewidth}
% \subfigure[]{\label{fig1:subfig:b}
% \includegraphics[width=0.45\linewidth]{fig/all_P.pdf}}
% \hspace{-1mm}
% %\hspace{0.01\linewidth}
% \centering

% \caption{ The confusion matrix (a) ResNet56+Each-based method, (b) ResNet56+ALL-based method.}
% \label{sample1}
% \end{figure}
\begin{table}[t]
\renewcommand\arraystretch{1.3}
\centering
\caption{The accuracy under different exchange times}
\label{raw}
\setlength{\tabcolsep}{2mm}{
\begin{tabular}{c|ccc}
\hline\hline
 \diagbox{Time of exchange}{Sample ratio}&{0.002} & {0.01}&{0.05}\\ \hline
0  & 0.1287  & 0.2200 &0.4506 \\ \hline
2  & 0.1588  & 0.3226 &0.5495 \\ \hline
4  & 0.2045  & 0.4099 &0.6064 \\ \hline
6  & \pmb{0.2239}  & \pmb{0.4638} &\pmb{0.6270} \\ 
\hline\hline
\end{tabular}}
\end{table}

\begin{table*}[!t]
\renewcommand\arraystretch{1.3}
\centering
\caption{The accuracy of the augmentation methods}
\label{flip}
\setlength{\tabcolsep}{2.4mm}{
\begin{tabular}{c|ccc|ccc|ccc}
\hline\hline
{Sample ratio}&\multicolumn{3}{c|}{0.002}&\multicolumn{3}{c|}{0.01} &\multicolumn{3}{c}{0.05}\\ \hline 
 \diagbox{Time of exchange}{Method} &$\rm Flip_{I}$&$\rm Flip_{IQ}$&$\rm Flip_{ALL}$ 
&$\rm Flip_{I}$&$\rm Flip_{IQ}$&$\rm Flip_{ALL}$ &$\rm Flip_{I}$&$\rm Flip_{IQ}$&$\rm Flip_{ALL}$\\ \hline

0    & 0.1642 &0.1800 &0.1876 &0.2817 &0.3343 &0.4131  & 0.5443 & 0.5581 &0.5752 \\ \hline 
2   & 0.2175 &0.2460 &0.2561 &0.4248 &0.4922 &0.5104   & 0.63403 & 0.6412  &\pmb{0.6462}\\ \hline 
4   & 0.2721 &0.2850 &0.3169  &0.4988 &\pmb{0.5275} &\pmb{0.5250}  & 0.6394 & \pmb{0.6458}  &0.6422\\ \hline 
6   & \pmb{0.2928} & \pmb{0.3511} &\pmb{0.3572} &\pmb{0.5215} &0.5256 &0.5204   & \pmb{0.6456}  & 0.6395  &0.6426\\
\hline\hline
\end{tabular}}
\end{table*}
\subsubsection{Augmentation in Few-Shot Scenario} 
We now evaluate the performance of our proposed methods in few-shot scenarios. We define the sample ratio as the ratio of the number of samples of each modulation category in the selected training set to the total number of samples of each modulation category in the original training set. %Take 0.2\%, 1\% and 5\% samples of each modulation category under each SNR as the received signal, i.e., sample ratio = 0.002, 0.01 and 0.05. The raw smaples $\mathcal{Y}$ received by the $C$ antennas or flipped datasets are randomly exchanged to obtain augmented data. At the same time, in order to explore the influence of exchange times on the accuracy of modulation recognition, we conducted different exchange times on the raw samples $\mathcal{Y}$ and flipped dataset $\mathcal{Y}_{\operatorname{flipI}}$, $\mathcal{Y}_{\operatorname{flipIQ}}$, and $\mathcal{Y}_{\operatorname{flipALL}}$. Finally, we use these augmented datasets to train MAMRnet to improve the recognition performance.
A 4-antenna receiving system is considered. To investigate the impact of the number of data exchanges on the effectiveness of our proposed data augmentation method, we vary the number of exchanges $D$, and evaluate the performance of MAMR-IQ method for $D = 2$, 4, and 6.  We first evaluate our proposed augmentation method without flipping and the results are presented in Table \ref{raw}. It is found that the recognition accuracy of the augmented dataset gradually increased as the number of data exchanges increased. For instance, when the sample ratio is 0.01 and the number of exchanges increases from 0 to 6, the recognition accuracy increases from 0.2200 to 0.4638. At  sample ratio of 0.01, the maximum improvement in performance is 24\%, while at sample ratios of 0.002 and 0.05, the corresponding improvements are 10\% and 17\%, respectively.  We also evaluate our proposed augmentation method combining with flipping and the results are shown in Table \ref{flip}. The proposed method of exchanging antennas also enhances the performance of the flipped method. When the sample ratio is 0.01 and the number of exchanges increases from 0 to 6, the $\rm Flip_{I}$-based MAMR-IQ method, $\rm Flip_{IQ}$-based MAMR-IQ method, and $\rm Flip_{ALL}$-based MAMR-IQ method show improved performance gains of 0.24, 0.19, and 0.11, respectively. This further validate the effectiveness of our proposed augmentation method.%Notably, the largest performance gain occurs when the number of exchanges is minimal.

\begin{figure}[!t]
\centering
\subfloat[\label{fig:Exchange}Pure Exchange ]{\includegraphics[width=.8\linewidth]{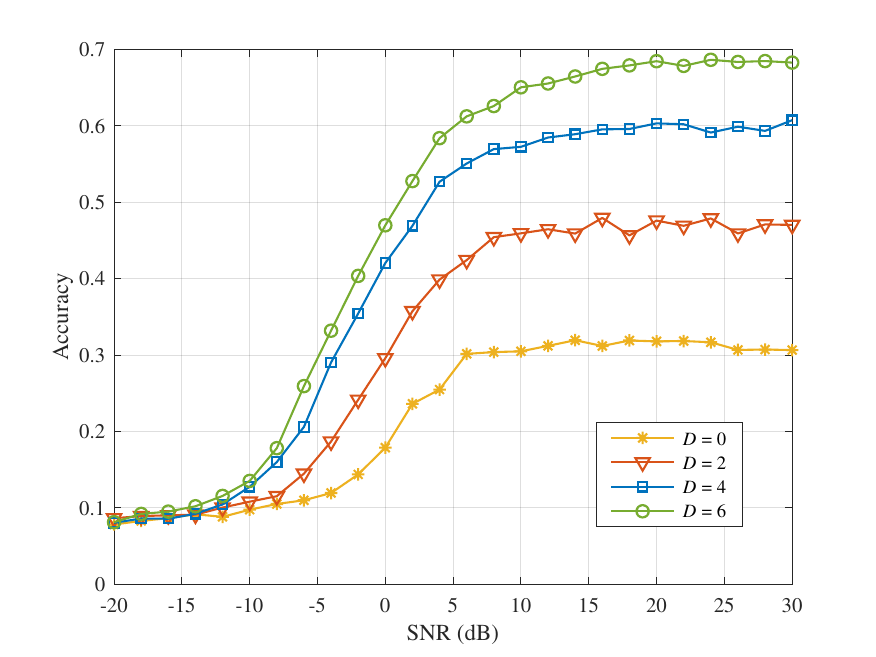}}%
\quad
\subfloat[\label{fig:Exchange1} $Flip_{I}$+Exchange method]{\includegraphics[width=.8\linewidth]{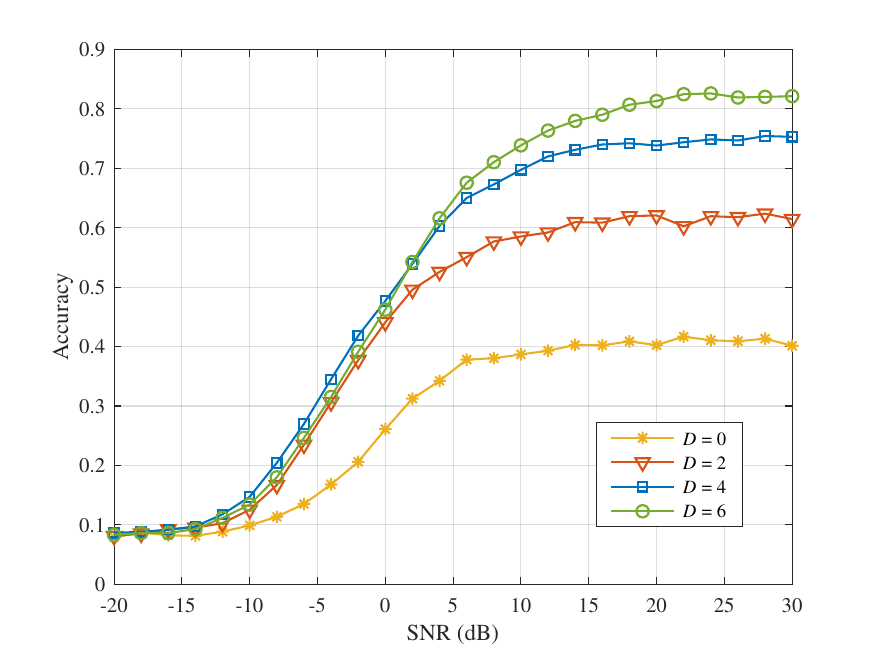}}\\
\caption{Recognition accuracy under different exchange times.}
\label{ex_0.01}
\end{figure}

\begin{figure*}[!ht]
\centering
\subfloat[\label{fig:sample0.01}raw samples and $D$ = 0]{\includegraphics[width=.25\linewidth]{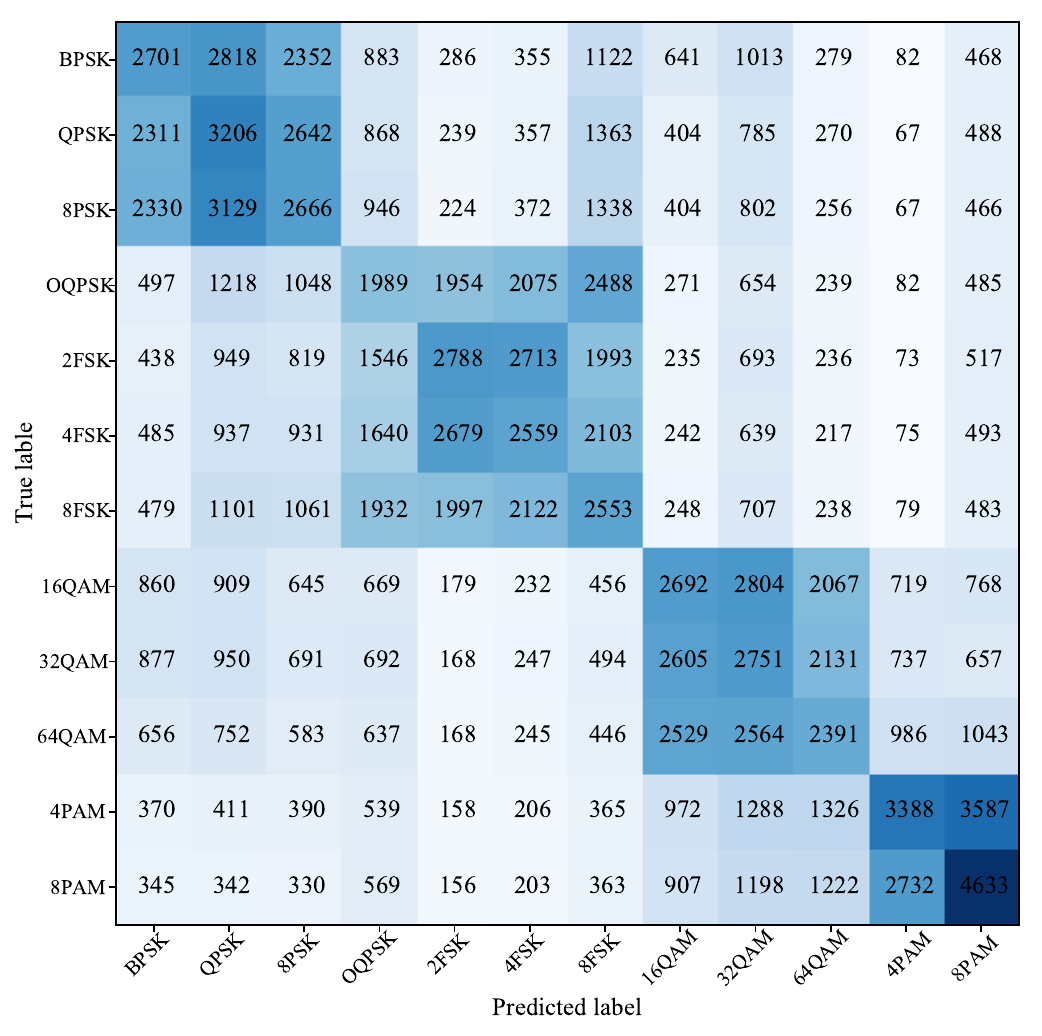}}%
\subfloat[\label{fig:exchange3}raw samples and $D$ = 2]{\includegraphics[width=.25\linewidth]{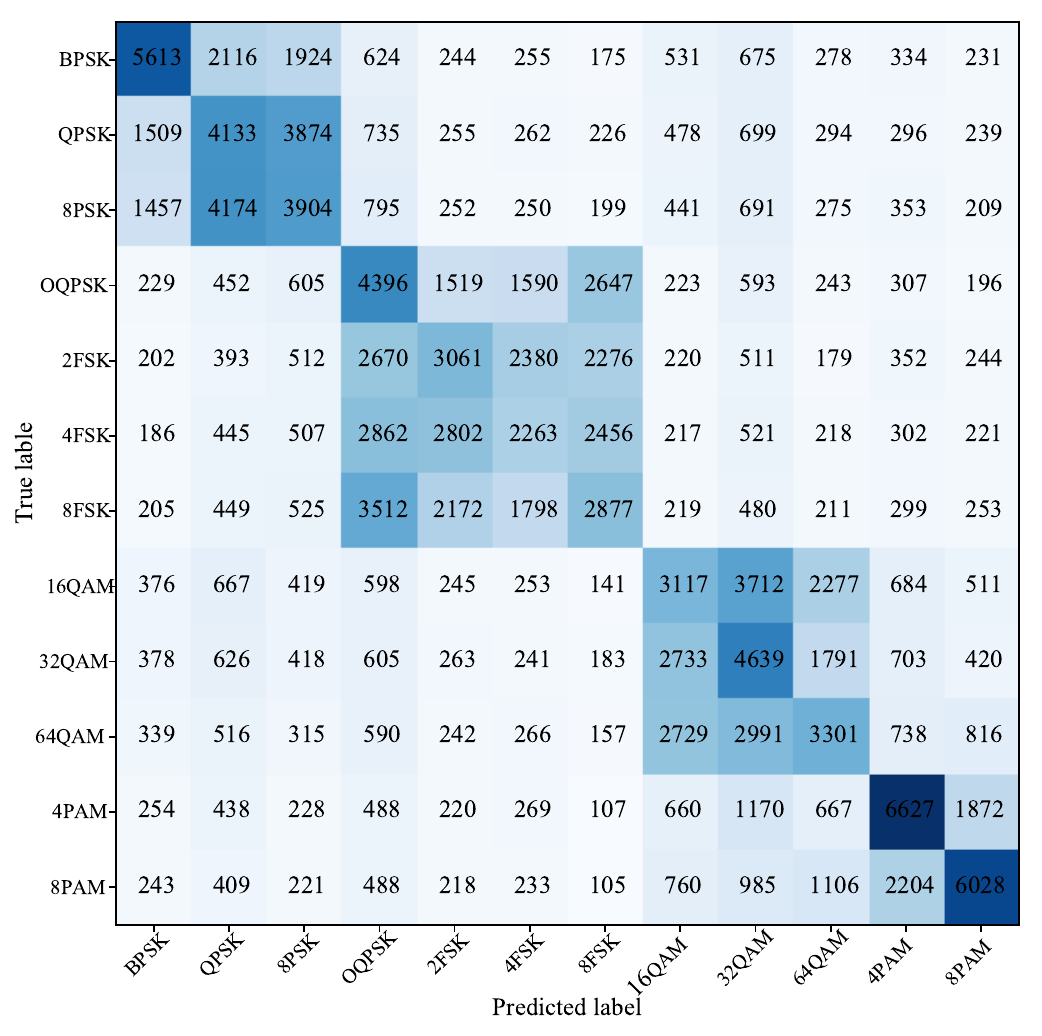}}
\subfloat[\label{fig:exchange5}raw samples and $D$ = 4]{\includegraphics[width=.25\linewidth]{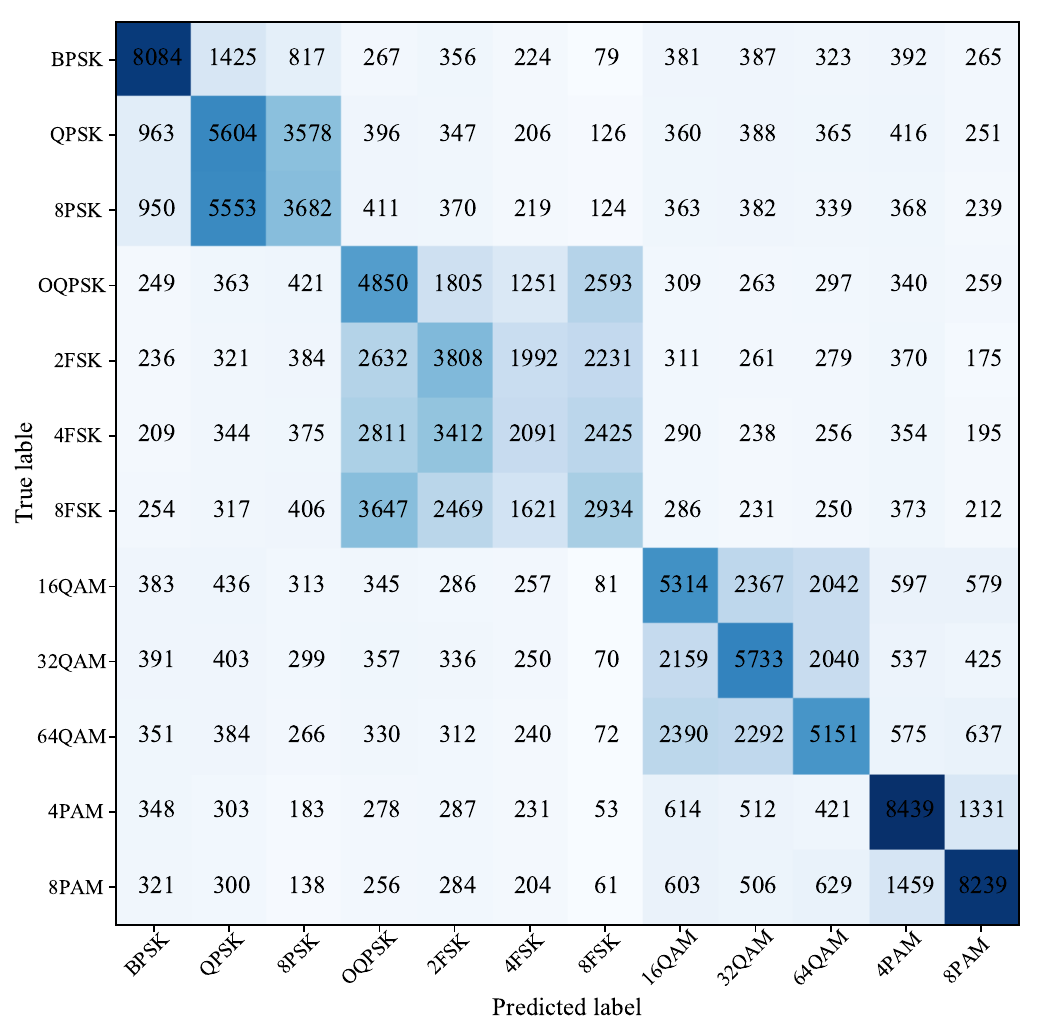}}%
\subfloat[\label{fig:exchange7}raw samples and $D$ = 6]{\includegraphics[width=.25\linewidth]{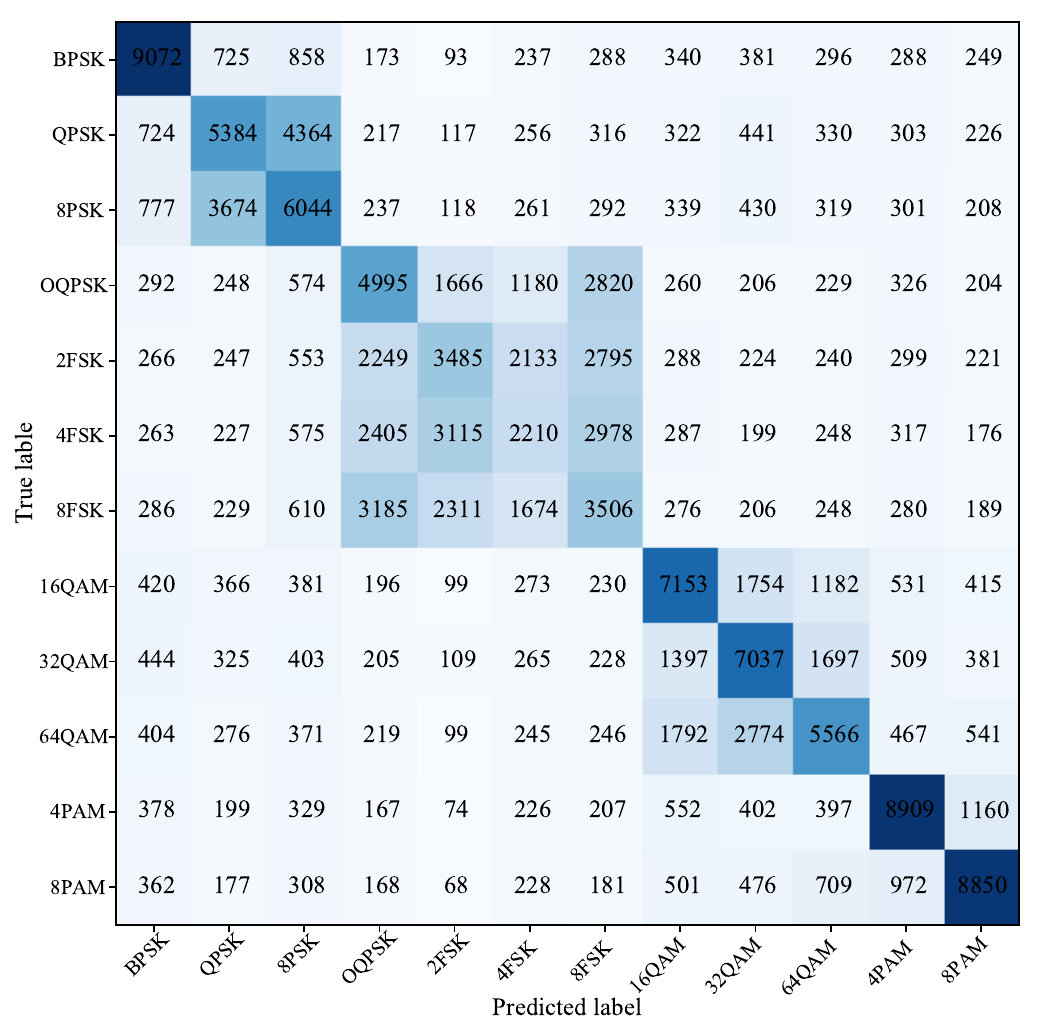}}%

\subfloat[\label{fig:flipI0.01} flipI samples and $D$ = 0]{\includegraphics[width=.25\linewidth]{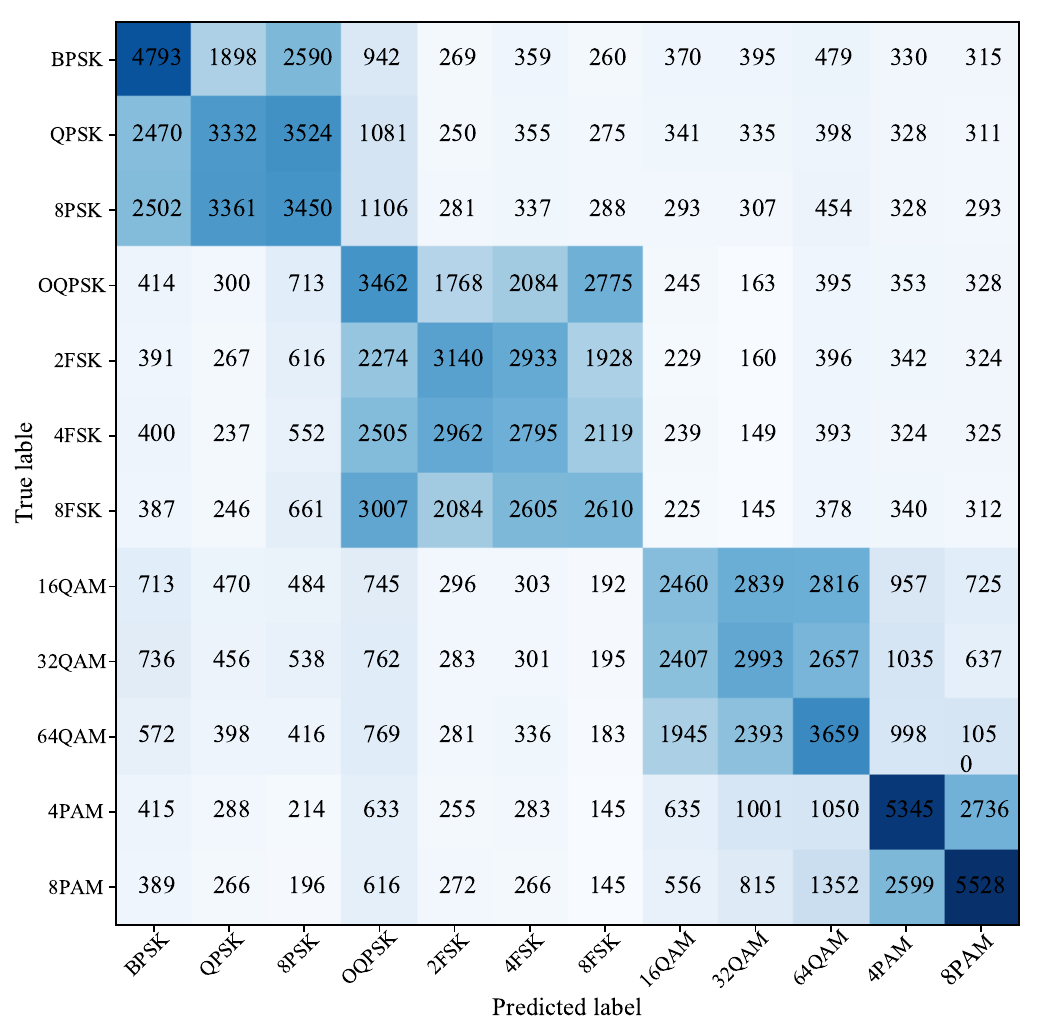}}%
\subfloat[\label{fig:flipI_EX2_0.01} flipI samples and $D$ = 2]{\includegraphics[width=.25\linewidth]{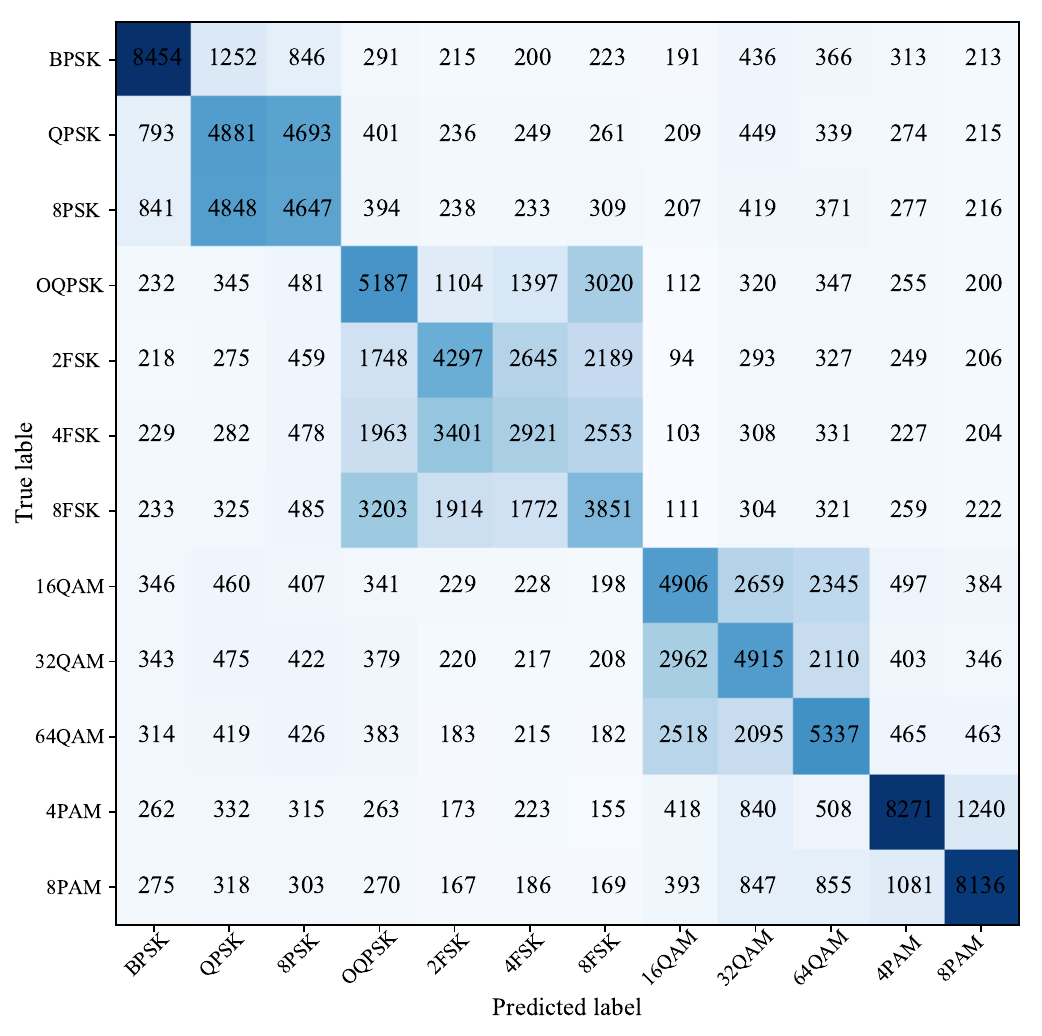}}
\subfloat[\label{fig:flipI_EX4_0.01} flipI samples and $D$ = 4]{\includegraphics[width=.25\linewidth]{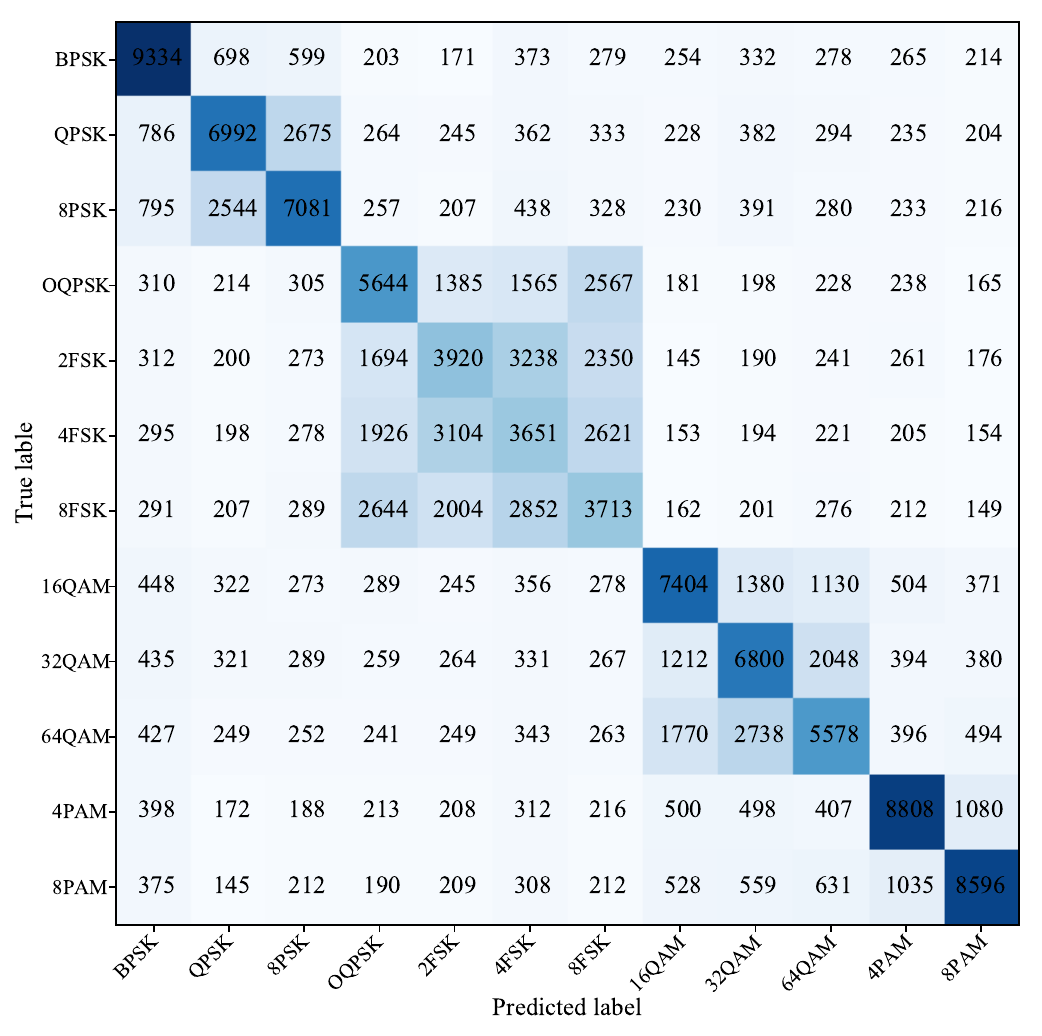}}%
\subfloat[\label{fig:flipI_EX6_0.01}  flipI samples and $D$ = 6]{\includegraphics[width=.25\linewidth]{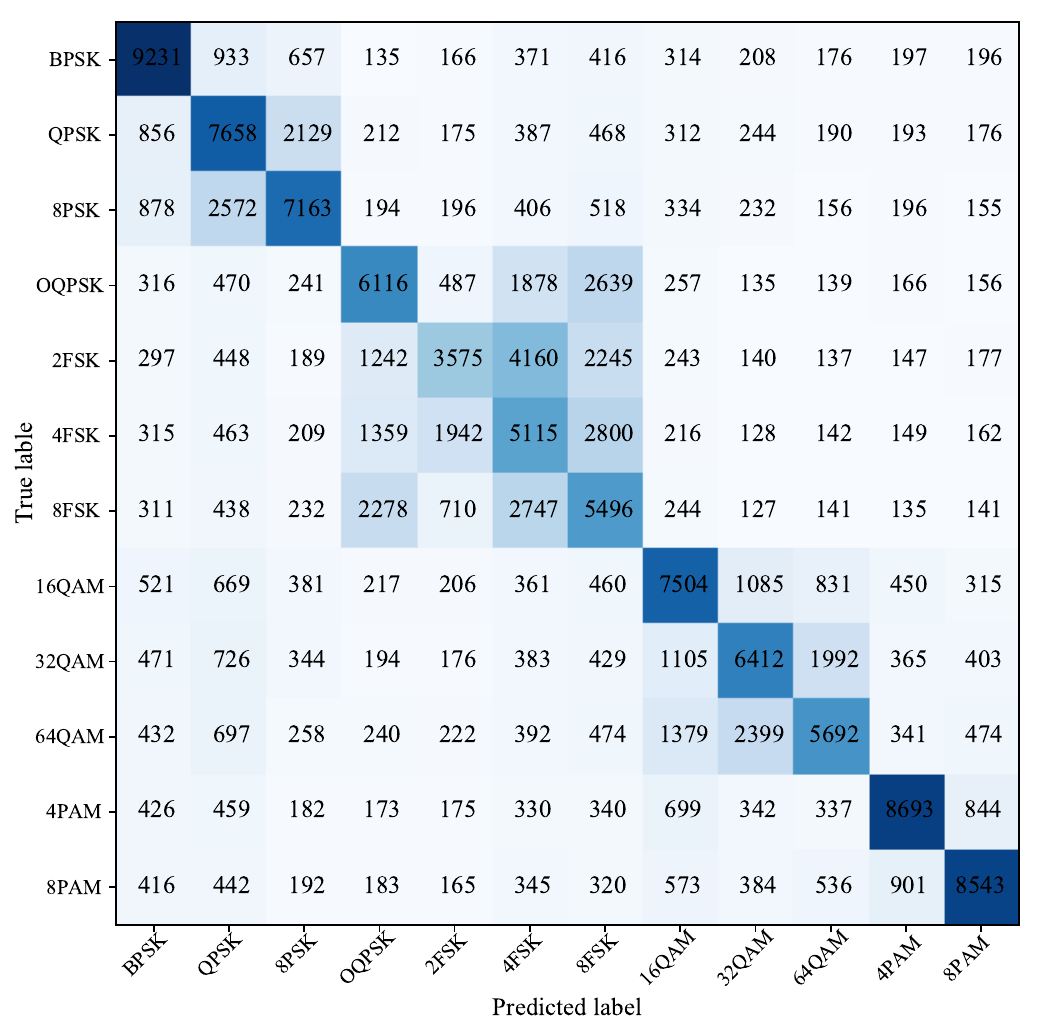}}%
\caption{The confusion matrices of augmentation method at sample ratio = 0.01.}
\label{exchange357}
\end{figure*}

We draw recognition accuracy under different exchange times of raw samples $\mathcal{Y}$ and the horizontally flipped samples $\mathcal{Y}_{\operatorname{flipI}}$ at sample ratio $=$ 0.01 in Fig. \ref{ex_0.01}. The results depicted  demonstrate that the proposed augmentation method leads to improved accuracy and the improvement is more obvious with medium to high SNR. Specifically, the augmentation method with raw IQ shows improvement in the SNR range of $-10$ dB to 30 dB, while the augmentation 
with horizontally flipping yields  improvement from SNR range of 5 dB to 30 dB. Notably, when $D$ is set to 6 and the SNR is 30 dB, the accuracy achieved by the augmented IQ samples can reach 0.7 while the accuracy without exchange is only 0.3. Similarly, the augmentation method with horizontally flipping further improve the accuracy from 0.4 to 0.82. Furthermore, we draw the confusion matrix under different exchange times in Fig. \ref{exchange357}. We can conclude that our proposed augmentation method without flipping has a significant improvement in the modulation types of BPSK, 16QAM, and 4PAM.
When the exchange times $D$ is 6, the number of correct identification of increases by 6300, 6600, and 6600 out of 13000 total training samples for BPSK, 16QAM, and 4PAM, respectively. For the horizontally flipped samples, the significant improvement occur with the modulation types of BPSK, QPSK, and 16QAM, and the number of correct identifications increased is 4700, 4300, and 5000 out of 13000 total training samples.

% \begin{figure*}
% \centering
% \subfigure[]{\label{fig:subfig3:sample0.01}
% \includegraphics[width=0.23\linewidth]{fig/SAMPLE_0.01.pdf}}
% \subfigure[]{\label{fig:subfig3:exchange3}
% \includegraphics[width=0.23\linewidth]{fig/Exchange2_0.01.pdf}}
% \subfigure[]{\label{fig:subfig3:exchange5}
% \includegraphics[width=0.23\linewidth]{fig/Exchange4_0.01.pdf}}
% \subfigure[]{\label{fig:subfig3:exchange7}
% \includegraphics[width=0.23\linewidth]{fig/Exchange6_0.01.pdf}}
% \centering
% \end{figure*}

% \begin{figure*}
% \centering
% \subfigure[]{\label{fig:subfig3:flipI0.01}
% \includegraphics[width=0.23\linewidth]{fig/flipI_0.01.pdf}}
% \subfigure[]{\label{fig:subfig3:flipI_EX2_0.01}
% \includegraphics[width=0.23\linewidth]{fig/flipI_EX2_0.01.pdf}}
% \subfigure[]{\label{fig:subfig3:flipI_EX4_0.01}
% \includegraphics[width=0.23\linewidth]{fig/flipI_EX4_0.01.pdf}}
% \subfigure[]{\label{fig:subfig3:flipI_EX6_0.01}
% \includegraphics[width=0.23\linewidth]{fig/flipI_EX6_0.01.pdf}}
% \centering

% \caption{The confusion matrices of augmentation method at sample ratio = 0.01. (a) raw samples and $D$ = 0; (b) raw samples and $D$ = 2; (c) raw samples and $D$ = 4; (d) raw samples and $D$ = 6 ; (e) the horizontally flipped samples and $D$ = 0;  (f) the horizontally flipped samples and $D$ = 2; (g) the horizontally flipped samples and $D$ = 4; (h) the horizontally flipped samples and $D$ = 6.}
% \label{exchange357}
% \end{figure*}

\subsection{ Complexity Analysis }
From the above experimental results, it can be seen that our proposed MAMR-IQ method has surprising performance gains compared with existing methods. In this section, we compare the time complexity and space complexity of these methods, i.e., computational complexity (FLOPs) and storage complexity (Parameters). Since the training process is carried out offline, we only evaluate the complexity of the inference process. The used network is ResNet56.

\subsubsection{Time Complexity}
Time complexity of MAMR-IQ method refers to the complexity of the ResNet56 network computation in the inference stage, whereas MAMR-DV method's complexity includes both the neural network computation and the direct voting process. Meanwhile, MAMR-WA method's complexity comprises the neural network computation and the weight averaging process. In the ResNet56 network, convolution operations take up a large amount of multiplication computation, so we ignore the operations that require less computation, such as ReLU, batch normalization and the operation of the fully connected layer. Similarly, the multiplication calculation in DV and WA operations can also be ignored. Therefore, We only consider multiplication calculation in convolution operation as the approximation of the total computational complexity. Therefore, the time complexity of ResNet56 $\mathcal{T}_{\rm ResNet}$ consists of the computational complexity $\mathcal{T}_{\rm Con1 }$ of the first convolution layer, and the computational complexity of the three residual blocks with 9 repetition times, i.e., $\mathcal{T}_{\rm 9Block1}$, $\mathcal{T}_{\rm 9Block2}$, and $\mathcal{T}_{\rm 9Block3}$ which are as follows:
\begin{equation}
\label{CON1}
   \begin{gathered}
\mathcal{T}_{\rm Con1} =F_l F_w K_l K_w T_{in} T_{out} 
=480C F_l F_w ,
\end{gathered}
\end{equation}
\begin{equation}
\label{block1}
   \begin{gathered}
\mathcal{T}_{\rm 9Block1} =13824F_l F_w,
\end{gathered}
\end{equation}
\begin{equation}
\label{block2}
   \begin{gathered}
\mathcal{T}_{\rm 9Block2} =26880F_l F_w,
\end{gathered}
\end{equation}
\begin{equation}
\label{block3}
   \begin{gathered}
\mathcal{T}_{\rm 9Block3} =53760F_l F_w,
\end{gathered}
\end{equation}
\begin{equation}
\begin{split}
\label{CN1}
\mathcal{T}_{\rm ResNet56}=\mathcal{T}_{\rm Con1}+ \mathcal{T}_{\rm 9Block1}+ \mathcal{T}_{\rm 9Block2}+
\mathcal{T}_{\rm 9Block3} \\
=480C F_l F_w+94464F_l F_w,
\end{split}
\end{equation}
The time complexity of MAMR-DV, MAMR-WA and MAMR-IQ are
% \begin{equation}
% \begin{split}
% \label{single}
% \mathcal{T}_{\rm MAMR-Single}=\mathcal{T}_{\rm ResNet56|C=1} 
% =480 F_l F_w+94464F_l F_w,
% \end{split}
% \end{equation}
\begin{equation}
\begin{split}
\label{DV}
\mathcal{T}_{\rm MAMR-DV}=C*\mathcal{T}_{\rm ResNet56|C=1} \\
=C(480 F_l F_w+94464F_l F_w),
\end{split}
\end{equation}
\begin{equation}
\begin{split}
\label{WA}
\mathcal{T}_{\operatorname{MAMR-WA}}=C*\mathcal{T}_{\rm ResNet56|C=1} \\
=C(480 F_l F_w+94464F_l F_w),
\end{split}
\end{equation}
\begin{equation}
\begin{split}
\label{ResNet56-ALL}
\mathcal{T}_{\rm MAMR-IQ}=\mathcal{T}_{\rm ResNet56}\\
=480C F_l F_w+94464F_l F_w,
\end{split}
\end{equation}
where $F_l$ * $F_w$ is the size of the feature map of the convolution output, $K_l$ * $K_w$ is the convolution kernel size, $T_{in}$ and $T_{out}$ are the numbers of input channels and output channels of the convolution kernel. 
%It is obvious that the time complexity of MAMR-IQ method proposed by us is much lower than that of MAMR-DV method and MAMR-WA method.

\subsubsection{Space Complexity}
The space complexity $\mathcal{S}$ is independent of the input size, but is related to the neural network model. Therefore, the space complexity mainly includes two parts. The first part $\mathcal{W}$ is the total number of weight parameters of the model, i.e., the model volume, and the other part $\mathcal{M}$ is the size of the feature map output by each layer of the network. In the actual inference process, since the output feature map of the current layer will cover the input feature map, the feature map of the space complexity is twice the maximum feature map of the model. Therefore, the expression of the feature map is
\begin{equation}
\begin{split}
\label{fea}
\mathcal{M}=2\mathcal{M_{\rm max}}=32F_l F_w.
\end{split}
\end{equation}

The total weight parameters of the model $\mathcal{W}_{\rm ResNet}$ consists of the parameters of the first convolution layer (denoted as $\mathcal{W}_{Con1}$), and the parameters of the three residual blocks with 9 repetition times, (denoted as $\mathcal{W}_{\rm 9Block1}$, $\mathcal{W}_{\rm 9Block2}$, and $\mathcal{W}_{\rm 9Block3}$, respectively).
\begin{equation}
\label{wCON1}
   \begin{gathered}
\mathcal{W}_{\rm Con1} = K_l K_w T_{in} T_{out}
=480C ,
\end{gathered}
\end{equation}
\begin{equation}
\label{wblock1}
   \begin{gathered}
\mathcal{W}_{\rm 9Block1} =13824,
\end{gathered}
\end{equation}
\begin{equation}
\label{wblock2}
   \begin{gathered}
\mathcal{W}_{\rm 9Block2} =53760,
\end{gathered}
\end{equation}
\begin{equation}
\label{wblock3}
   \begin{gathered}
\mathcal{W}_{\rm 9Block3} =215040,
\end{gathered}
\end{equation}
\begin{equation}
\begin{split}
\label{wCN1}
\mathcal{W}_{\rm ResNet56}=\mathcal{W}_{\rm Con1}+ \mathcal{W}_{\rm 9Block1}+ \mathcal{W}_{\rm 9Block2}\\+
\mathcal{W}_{\rm 9Block3} =480C +282624,
\end{split}
\end{equation}
Therefore, the space complexity of these methods is
% \begin{equation}
% \begin{split}
% \label{wCN11}
% \mathcal{W}_{\rm MAMR-Single}=\mathcal{W}_{\rm ResNet56|C=1} =283104+32F_l F_w,
% \end{split}
% \end{equation}
\begin{equation}
\begin{split}
\label{wDV}
\mathcal{W}_{\rm MAMR-DV}=C*\mathcal{W}_{\rm ResNet56|C=1} + \mathcal{M} \\
=C(480 +282624) +32F_l F_w,
\end{split}
\end{equation}
\begin{equation}
\begin{split}
\label{wWA}
\mathcal{W}_{\rm MAMR-WA}=C*\mathcal{W}_{\rm ResNet56|C=1}+\mathcal{M}
 \\
=C(480 +282624)+32F_l F_w,
\end{split}
\end{equation}
\begin{equation}
\begin{split}
\label{wResNet56-s}
\mathcal{W}_{\rm MAMR-IQ}=\mathcal{W}_{\rm ResNet56}+\mathcal{M}\\
=480C +282624+32F_l F_w,
\end{split}
\end{equation}
%As the conclusion of the time complexity, the space complexity of MAMR-IQ method proposed by us is less than that of MAMR-DV method and MAMR-WA method.
In order to facilitate a more intuitive comparison of the complexity of these methods, we have set the number of antennas to 4 and the corresponding results are presented in Table .\ref{csr_table_fuzadu}. As can be observed from the table, our proposed MAMR-IQ method's complexity is only slightly higher than that of MAMR-Single method. In contrast, the complexity of MAMR-DV and MAMR-WA methods is nearly four times higher than that of MAMR-Single method. This indicates that our proposed MAMR-IQ obtains higher recognition accuracy with less complexity over existing MAMR-DV and MAMR-WA methods.

\begin{table}[!t]
\renewcommand\arraystretch{1.3}
\centering
\caption{FLOPs and Params of Different Mehtods}
\setlength{\tabcolsep}{2.5mm}{
\begin{tabular}{c|ccc}
\hline\hline
 \diagbox{Methods}{Complexity}&{FLOPs (M)} & {Params (M)} \\ \hline
MAMR-Single  & \bf 24.73  & \bf 0.288  \\ \hline
MAMR-DV  & 98.92  & 1.144 \\ \hline
MAMR-WA  & 98.92  & 1.144  \\ \hline
MAMR-IQ  & 25.09  & 0.289   \\ 
\hline\hline
\end{tabular}}
\label{csr_table_fuzadu}
\end{table}

%----------------------------------------------------------------------------------------
\section{Conclusion}
%We propose a MAMR-IQ method which concatenates the raw IQ components of signals received by multiple antennas and then feeds them to our designed neural network to learn the hidden features and thus identify the modulation types.
We have utilized deep learning in a multi-antenna receiving system for performing modulation recognition tasks. Specifically, we have concatenated the IQ sequences obtained by multiple antennas and feed them into neural network for modulation recognition. To showcase the effectiveness of our proposed approach, we have compared it against the existing MAMR-DV and MAMR-WA methods. The simulation results have indicated that our proposed MAMR-IQ method outperforms MAMR-DV and MAMR-WA methods both in terms of accuracy and complexity. We have also proposed an augmentation method in few-shot scenarios, which involves generating augmented samples by exchanging the IQ sequences received by any two antennas. Simulation results have shown that our proposed augmentation method can improve the recognition accuracy in few-shot scenarios.% In the future work, we will further study the multi-antenna receiving system and fully combine the deep learning and other existing augmentation method to deal with other radio problems.

%----------------------------------------------------------------------------------------
\section*{ACKNOWLEDGEMENT}
\label{ACKNOWLEDGEMENT}
This work was supported by Natural Science Foundation of Zhejiang province LQN26F010033.

\small
\bibliographystyle{ieeetr}
\bibliography{citationlist}

\end{document}